\documentclass[12pt]{article}

\usepackage{amssymb,amsmath}	% Mathematical symbols and functions
\usepackage{graphicx}			% Figures
\usepackage{subfigure}			% Sub figure system
\usepackage{color}				% color text
\usepackage{authblk}	 		% Title page
\usepackage{hyperref}			% Links of references
\usepackage{cite}
\begin{document}

\baselineskip 6mm
\renewcommand{\thefootnote}{\fnsymbol{footnote}}

%------------ Chanyong Park's macro's, etc  -----------

\newcommand{\nc}{\newcommand}
\newcommand{\rnc}{\renewcommand}

%%%%%%%%%%%%%%%%%%%%%% Equation Numbering %%%%%%%%%%%%%%%%%%%%%%%
%\makeatletter \rnc{\theequation}{\thesection.\arabic{equation}}
%\@addtoreset{equation}{section} \makeatother

%%%%%%%%%%%%%%%%%%%%%%%%%%%%%%%%%%%%%%%%%%%%%%%%%%%%%%%%%%%%%%%%%
%                                                               %
%                NEW COMMANDS AND MACROS                        %
%                                                               %
%%%%%%%%%%%%%%%%%%%%%%%%%%%%%%%%%%%%%%%%%%%%%%%%%%%%%%%%%%%%%%%%%

\newcommand{\tcb}{\textcolor{blue}}
\newcommand{\tcr}{\textcolor{red}}
\newcommand{\tcg}{\textcolor{green}}

%%%%% Simplify some frequently used LaTeX commands %%%%%

\def\ba{\begin{array}}
\def\ea{\end{array}}
\def\be{\begin{eqnarray}}
\def\ee{\end{eqnarray}}
\def\nn{\nonumber\\}
\newcommand{\bea}{\begin{eqnarray}}
\newcommand{\eea}{\end{eqnarray}}

%%%%%  Temporary notation %%%%

\def\ct{\cite}
\def\la{\label}
\def\eq#1{(\ref{#1})}

%%% Greek letters %%%

\def\a{\alpha}
\def\b{\beta}
\def\g{\gamma}
\def\G{\Gamma}
\def\d{\delta}
\def\D{\Delta}
\def\e{\epsilon}
\def\et{\eta}
\def\ph{\phi}
\def\Ph{\Phi}
\def\ps{\psi}
\def\Ps{\Psi}
\def\k{\kappa}
\def\l{\lambda}
\def\L{\Lambda}
\def\m{\mu}
\def\n{\nu}
\def\th{\theta}
\def\Th{\Theta}
\def\r{\rho}
\def\s{\sigma}
\def\S{\Sigma}
\def\ta{\tau}
\def\o{\omega}
\def\O{\Omega}
\def\pr{\prime}
\def\nn{\nonumber}
\def\pa{\partial}
\def\la{\label}

%%%%% Mathematical Symbols

\def\half{\frac{1}{2}}

\def\goto{\rightarrow}

\def\na{\nabla}
\def\grad{\nabla}
\def\curl{\nabla\times}
\def\div{\nabla\cdot}
\def\pa{\partial}
\def\fr{\frac}

\def\bra{\left\langle}
\def\ket{\right\rangle}
\def\lb{\left[}
\def\lc{\left\{}
\def\ls{\left(}
\def\lp{\left.}
\def\rp{\right.}
\def\rb{\right]}
\def\rc{\right\}}
\def\rs{\right)}

\def\vac#1{\mid #1 \rangle}

%%%%  Special symbol

\def\td#1{\tilde{#1}}
\def\check{ \maltese {\bf Check!}}

%%%%% Roman pont in math

\def\Tr{{\rm Tr}\,}
\def\det{{\rm det}}

%%%%% Special format

\def\bc#1{\nnindent {\bf $\bullet$ #1} \\ }
\def\ch {$<Check!>$ }
\def\ss {\vspace{1.5cm}}
\def\text#1{{\rm #1}}
\def\Id{\mathds{1}}

\begin{titlepage}

%---------------- preprint number ---------------
%\hfill\parbox{5cm} { }

\vspace{25mm}

\begin{center}
%------------------------ title ------------------------
{\Large \bf c-theorem of the entanglement entropy  }

%---------------- authors and addresses ----------------
\vskip 1. cm
  {Chanyong Park$^{a,b,c}$\footnote{e-mail : cyong21@gist.ac.kr} Daeho Ro$^{b}$\footnote{e-mail : daeho.ro@apctp.org} and Jung Hun Lee$^{a,b}$\footnote{e-mail : junghun.lee@gist.ac.kr}}

\vskip 0.5cm

{\it $^a\,$ Department of Physics and Photon Science, Gwangju Institute of Science and Technology,  Gwangju, 61005, Korea }\\
{\it $^b\,$Asia Pacific Center for Theoretical Physics, Pohang, 37673, Korea}\\
{\it $^c\,$ Department of Physics, Postech, Pohang 37673, Korea } \\
\end{center}

\thispagestyle{empty}

\vskip2cm

%----------------------- abstract ----------------------

\centerline{\bf ABSTRACT} \vskip 4mm

\vspace{1cm}

We holographically investigate the renormalization group flow in a two-dimensional conformal field theory deformed by a relevant operator. If the relevant operator allows another fixed point, the UV conformal field theory smoothly flows to a new IR conformal field theory. From the holographic point of view, such a renormalization group flow can be realized as a dual geometry interpolating two different AdS boundaries. On this interpolating geometry, we investigate how the $c$-function of the entanglement entropy behaves along the RG flow analytically and numerically, which reproduces the expected central charges of UV and IR. We also show that the $c$-function monotonically decreases from UV to IR without any phase transition.

\vspace{2cm}
%PACS numbers:

%\today

\end{titlepage}

\renewcommand{\thefootnote}{\arabic{footnote}}
\setcounter{footnote}{0}

\newpage
\tableofcontents
\setcounter{page}{1}
%\tableofcontents

%%%%%%%%%%%%%%%%%%%%%%%%%
%                                                                            %
%   Sec.  Introduction                                           %
%                                                                            %
%%%%%%%%%%%%%%%%%%%%%%%%%

\section{Introduction\label{sec:1}}

Recently, considerable attention has been paid to calculate and understand the entanglement entropy and its cousins which are important concepts to figure out a variety of quantum features of a many-body system \cite{Calabrese:2004eu,Calabrese:2005zw,Calabrese:2009qy}. Despite their importance, it still remains a difficult task to calculate exactly the entanglement entropy of an interacting quantum field theory (QFT) we are interested in. In this situation, the AdS/CFT correspondence proposed in the string theory becomes a new and fascinating tool \cite{Maldacena:1997re,Gubser:1998bc,Witten:1998qj,Witten:1998zw}, because it enables us to calculate nonperturbatively the entanglement entropy of a strongly interacting quantum system \cite{Ryu:2006bv,Ryu:2006ef,Hubeny:2007xt,Casini:2011kv,Lewkowycz:2013nqa}. In this work, we investigate the renormalization group (RG) flow of the entanglement entropy and the change of the central charge along the RG flow when a two-dimensional conformal field theory (CFT) is deformed by a relevant operator \cite{Casini:2004bw,Casini:2006es,Albash:2011nq,Klebanov:2012yf,Cremonini:2013ipa,Faulkner:2014jva,Park:2014gja,Park:2015hcz,Casini:2015ffa,Kim:2016ayz,Kim:2016jwu,Bueno:2016rma,Kim:2016hig,Kim:2017lyx,Jang:2017gwd,Ghosh:2017big,Narayanan:2018ilr,Koh:2018rsw}.

The AdS/CFT correspondence proposed that the non-perturbative features of a conformal field theory (CFT) can be understood from a one-dimensional higher AdS geometry. Intriguingly, it was shown that the holographic techniques like a holographic renormalization and holographic entanglement entropy exactly reproduce the known results of a CFT \cite{Ryu:2006bv,Ryu:2006ef}. However, when a CFT is deformed by a relevant operator, the CFT description is not valid anymore because the relevant deformation breaks the conformal symmetry. Another important point we should note is that a relevant deformation can lead to a new CFT at an IR fixed point. In order to study holographically such a nontrivial RG flow, we first need to find a geometry realizing the nontrivial RG flow from a UV to IR fixed point. Due to the existence of the UV and IR fixed points, the dual geometry must interpolate two AdS spaces defined at the asymptotic boundary and at the center of a dual geometry. A dual geometry connecting two fixed points can be achieved by introducing a scalar field with an appropriate potential\cite{Skenderis:1999mm}. The gravity theory, which allows the smooth interpolation between UV and IR has been known not only in an $N=2$ gauged supergravity theory on $AdS_3$ but also other higher dimensional theories \cite{Deger:1999st,Deger:2002hv}. On deformed $AdS_3$ geometry interpolating UV and IR, in this work, we investigate how the dual field theory evolves along the RG flow by using the holographic renormalization and entanglement entropy.

Related to the RG flow, one of the important quantities we are interested in is the central charge representing the degrees of freedom of the field theory. When the field theory is deviated from a fixed point due to the deformation, in general, the central charge is not well defined. Even in this case, we can define a $c$-function which reduces to the central charge at a fixed point where the CFT naturally occurs. Interestingly, it has been claimed that the $c$-function monotonically decreases along the RG flow, which is called the $c$-theorem \cite{Zamolodchikov:1986gt,Cardy:1988cwa,Komargodski:2011vj,Komargodski:2011xv}. In this work, we investigate the entanglement entropy and the $c$-function from the dual interpolating geometry. Through this, we can get some nonperturbative information about the change of the degrees of freedom following the AdS/CFT prescription. In order to look into the $c$-function analytically, we apply a thin-wall approximation together with a natural junction condition requiring the continuity of the minimal surface at the wall. Instead of the natural junction condition, Ref. \cite{Myers:2012ed} utilized the reality of the entanglement entropy formula. The junction condition we exploit automatically satisfies the reality condition in \cite{Myers:2012ed} because the range of the variable determined by the junction condition is in the range derived from the reality. In general, the thin-wall approximation contains an inevitable error because the background geometry of the thin-wall approximation suddenly changes at the wall. In spite of this fact, the analytic analysis with the natural junction condition shows that the $c$-function continuously and monotonically decreases along the RG flow differently from the result of Ref. \cite{Myers:2012ed}. This implies the absence of a phase transition which results in breaking of the $c$-theorem. In order to check the thin-wall approximation we use, we also investigate the exact $c$-function numerically and show that the monotonic decreasing of the $c$-function in the thin-wall approximation is in agreement with the numerical result, as we expected.

The rest of this paper is organized as follows: In Sec. 2, we take into account an $N=2$ gauged supergravity theory and classify the possible geometric solutions. In a specific parameter range, the theory allows a geometric solution interpolating two AdS spaces which, on the dual field theory point of view, describes a nontrivial RG flow from a UV to IR fixed point. On this interpolating geometry, in Sec. 3, we study the RG flow of the entanglement entropy analytically with a thin-wall approximation. In particular, we concentrate on how the central charge representing the degrees of freedom is modified along the RG flow. We further compare the analytic result obtained by a thin-wall approximation with the exact and numeral result in Sec. 4. Finally, we finish this work with some concluding remarks in Sec. 5.

%%%%%%%%%%%%%%%%%%%%%%%%%
%                                                                            %
%   Sec.  Contents                                                %
%                                                                            %
%%%%%%%%%%%%%%%%%%%%%%%%%

%%%%%%%%%%%%%%%%%%%%%%%%%%%%%%%%%%
\section{$N=2$ gauged supergravity on $AdS_3$}
\label{sec:2}
%%%%%%%%%%%%%%%%%%%%%%%%%%%%%%%%%%

We take into account a dual field theory of an $N=2$ gauged supergravity on $AdS_3$ with supergravity and scalar multiplets. The supergravity multiplet is composed of a graviton $e^a_\m$, a complex gravitini $\psi_\m$ and a gauge field $A_\m$, while the scalar multiplet contains complex scalar fields $\Ph^a$ and fermions $\l^r$. Hereafter, we consider only one scalar multiplet for simplicity. Denoting the modulus of the complex scalar field as $\ph =|\Ph|$, the vacuum geometry of the gauged supergravity can be determined by this scalar field. The vacuum geometry appears as a solution of the following action \cite{Deger:1999st,Deger:2002hv}
\be		\la{eq:sugra-action}
S= \fr{1}{16 \pi G} \int d^3 x \sqrt{-g} \ls  {\cal R} - \fr{1}{a^2} \pa_\m \ph \pa^\m \ph - V(\ph) \rs,
\ee
with the scalar potential
\be		\la{eq:pot}
V(\ph) = 2 \L_{uv}  \cosh^2 \ph \lb (1 - 2 a^2) \cosh^2 \ph + 2 a^2 \rb ,
\ee
where $\L_{uv} = -1/R_{uv}^2$ indicates a negative cosmological constant. Note that, if we take the gravitational Newton constant to be $G=1/4 \pi$, the above action reduces to the one studied in \cite{Deger:2002hv}.

The vacuum structure of this model can be classified by the free parameter $a$ of the scalar potential. The scalar potential $V(\ph)$ always has a local extremum at $\ph=0$ with $V(0) = 2 \L_{uv}$. At this point, the geometric solution satisfying the Einstein equation leads to an AdS space with the AdS radius $R_{uv}$. Relying on the value of $a$, the local extremum has a different physical meaning:
\begin{itemize}

\item For $a^2 \ge 1$, the local extremum at $\ph=0$ becomes stable because it is a global minimum as shown in Fig.\,\ref{fig:pot}. Its dual QFT is mapped to a two-dimensional CFT with a central charge $c=3 R_{uv}/2G$.

\item For $1/2 < a^2 < 1$, the local extremum becomes a local maximum which is unstable. In addition, the scalar potential $V(\ph)$ allows another local minima at $\pm \ph_{ir}$,
\be		\label{res:irphi}
\ph_{ir} =  {\rm arccosh} \ls \fr{a}{\sqrt{2 a^2 -1} } \rs ,
\ee
with
\be
V(\ph_{ir})  =  - \fr{2}{R_{ir}^2 }  ,
\ee
where $V(\ph_{ir}) < V(0)$ and $R_{ir}= R_{uv} \sqrt{2 a^2 -1} /a^2$. From now on, we focus on the non-negative scalar field, $\ph \ge 0$. At the new local minimum, another $AdS_3$ geometry naturally appears as a new vacuum solution with a new $AdS$ radius $R_{ir}$ (see Fig.\,\ref{fig:pot}). Since the geometric solution at $\ph=0$ is unstable, we can expect a new solution interpolating these two different AdS geometries. From the dual QFT point of view, it corresponds to a deformation from a UV CFT with $\ph=0$ to another IR CFT with $\ph=\ph_{ir}$. In this case, the bulk scalar field is matched to a relevant deformation operator generating a nontrivial RG flow.

%%%%%%%%%%%%%%%%%%
\begin{figure}
\centering
\includegraphics[width=0.5\textwidth]{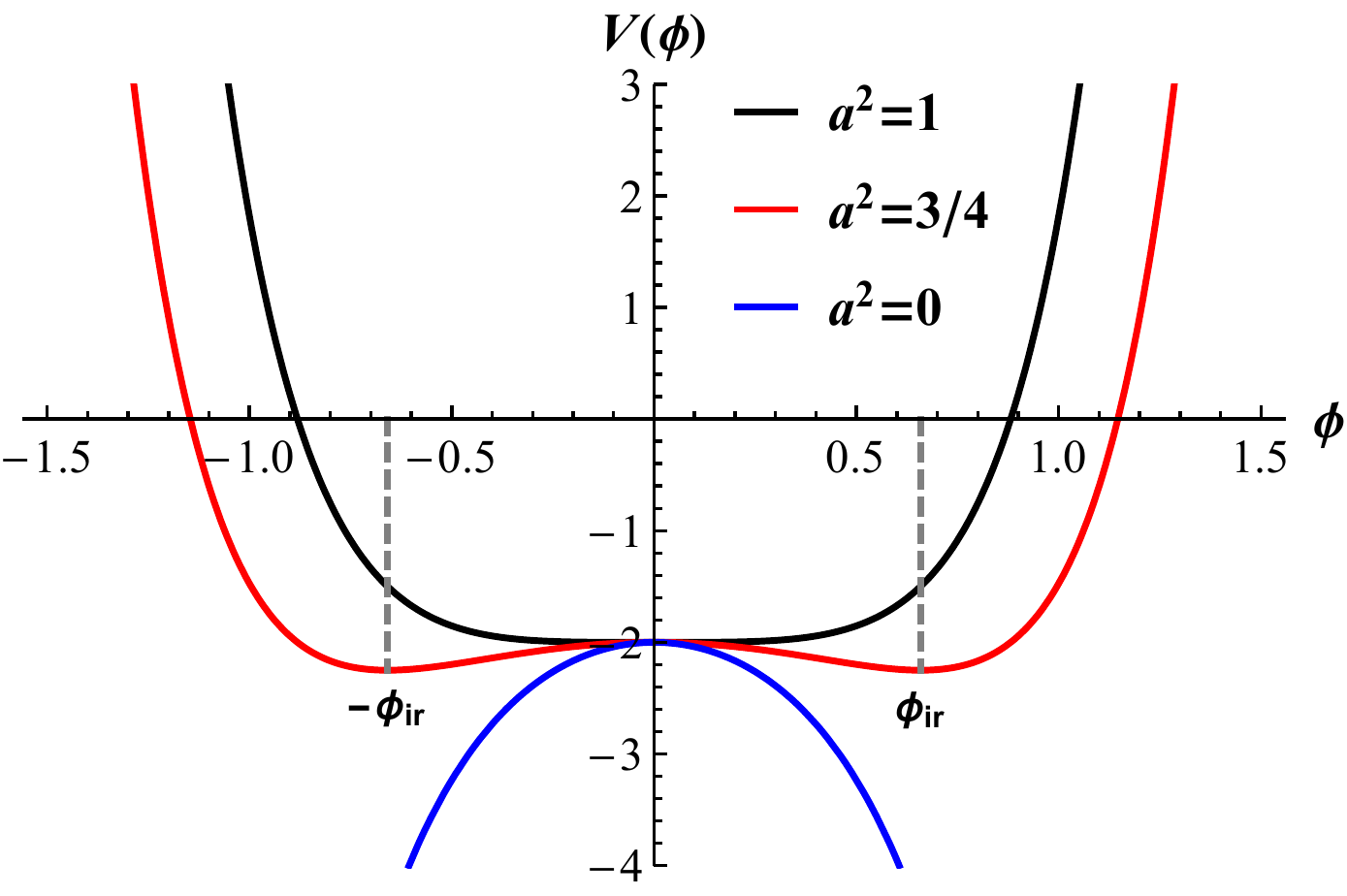}
\caption{Potential with respect to several values of $a^2$.}
\label{fig:pot}
\end{figure}
%%%%%%%%%%%%%%%%%%%

\item For $0< a^2 \le1/2$, the local extremum still remains as a local maximum but there is no additional local minimum. Although the scalar field fluctuation is also dual to a relevant deformation, it leads to a run-away potential and runs to $\ph =\infty$. Because of this fact, the IR geometry defined at $\ph =\infty$ becomes singular. On the dual field theory side, it indicates incompleteness of IR physics.
\end{itemize}

In order to figure out more details of the vacuum geometry, we introduce a normal coordinate
\be   	\la{metric:adsy}
ds^2 = e^{2 A(y)} \ls - dt^2 + dx^2 \rs + dy^2 ,
\ee
where the equations of motion derived from \eqref{eq:sugra-action} have relatively simple forms
\be		\la{eq:UVflucmetric}
\phi'' + 2 A' \phi' &=& \frac{a^2}{2} \frac{\partial V}{\partial \phi} , \nonumber \\
2 (A')^2 &=&  \frac{(\phi')^2}{a^2} - V  , \nonumber \\
A'' + 2 (A')^2 &=& - V .
\ee 
Here the prime means a derivative with respect to $y$.  Before studying the geometric solution satisfying these equations, there are several important points we must remember. First, since combining first two equations reproduces the last equation,  only two of them are independent. In the normal coordinate, second, an AdS geometry is represented by $\ph'=0$ and $A(y) = y/R$ with an AdS radius $R$. Lastly, the range of $y$ in the normal coordinate is extended to the entire region of $y$, $-\infty < y < \infty$. In this case the UV limit corresponds to $y \to \infty$, whereas the IR limit occurs at $y=- \infty$. 

At the local extremum with $\ph=0$, an AdS space with an AdS radius $R_{uv}$ satisfies the above equations and the unknown metric function is given by
\be 		\la{res:UVgeometry}
A  =  \fr{y}{R_{uv}} .
\ee
For $a^2 \ge 1$, since this solution represents the geometry of the global minimum, it is a stable solution. For $0< a^2 < 1$, however, the local extremum described by \eq{res:UVgeometry} corresponds to a local maximum which is unstable under a small perturbation. Turning on small fluctuations for $\ph$ and $A$ near $\ph=0$
\be			\la{ansatz:UVA}
\ph = 0 + \d\ph \quad {\rm and} \quad A = \fr{y}{R_{uv}} + \d A ,
\ee
the linearized equation of the scalar field fluctuation yields 
\be
0 = R_{uv}^2 \delta \phi '' +2 R_{uv} \delta \phi' -4 a^2 ( a^2-1)  \delta \phi  .
\ee
The solution for $a^2 \ne 1/2$ is given by
\be			\la{sol:sfluct}
\d\ph = c_1 e^{ - 2 (1 - a^2) y/R_{uv}}   \ls1 + \cdots \rs + c_2 e^{-2 a^2 y/R_{uv}} \ls1 + \cdots \rs  ,
\ee
where the ellipses indicate higher order corrections. Note that the first term in the above solution gives rise to a leading contribution for $1/2 <a^2 <1$, while the second term becomes dominant 
for $ 0<a^2 <1/2$. Substituting this scalar fluctuation solution into the second equation of \eq{eq:UVflucmetric}, we obtain the gravitational backreaction of the scalar field fluctuation 
\be			\la{res:uvflucA}
\d A = - \fr{  c_1^2}{4 a^2} \  e^{ - 4 (1 - a^2) y/R_{uv}}  \ls1 + \cdots \rs   -\frac{c_2^2 }{4 a^2 }  \ e^{- 4 a^2 y/R_{uv}}  \ls1 + \cdots \rs  .
\ee
For  $1/2 <a^2 <1$, the first term is dominant and, following the AdS/CFT correspondence, $c_1$ and $c_2$ can be interpreted as the source and vev of a dual operator, respectively. For $0<a^2 <1/2$, on the other hand, $c_1$ and $c_2$ exchange their roles. In other words, $c_2$ and $c_1$ play the role of the source and vev. For $a^2=1/2$, the two independent solutions in \eq{sol:sfluct} become degenerate. In this case, the solution of $\d\phi$ leads to
\be			\la{sol:uvscalarpert}
\d\phi= \bar{c}_1 e^{-y/R_{uv}} \ls1 + \cdots \rs  + \bar{c}_2 \fr{y}{R_{uv}} e^{-y/R_{uv}} \ls1 + \cdots \rs  .
\ee
The corresponding gravitational backreaction of the scalar fluctuation reads
\be			\la{res:a2half}		
\d A= -\frac{\bar{c}_1^2}{2}e^{-2y/R_{uv}} \ls1 + \cdots \rs  - \frac{\bar{c}_2^2}{2} \fr{y^2}{R_{uv}^2} e^{-2y/R_{uv}} \ls1 + \cdots \rs  .
\ee

For $1/2 <a^2 <1$, as mentioned before, there exists another local minimum with $\ph'=0$ at $\ph=\ph_{ir}$. This implies that another AdS geometry becomes a solution of the equations of motion at $\ph_{ir}$. In this case, the unknown function $A(y)$ is given by
\be		\la{con:irboundary}
A  \sim \fr{y}{R_{ir}} ,
\ee
where $R_{ir}= R_{uv} \sqrt{2 a^2 -1} /a^2$ is determined from the scalar potential\cite{Henningson:1998gx}. Near the new local minimum, the scalar fluctuation satisfying the linearized equation of motion becomes
\be			
\d\ph   = c_3 e^{- \D'_- y /R_{ir}} \ls1 + \cdots \rs  + c_4 e^{- \D'_+ y/R_{ir}}   \ls1 + \cdots \rs  ,
\ee
with
\be
\D'_{\pm} = 1 \pm \sqrt{9 - 8 a^2} .
\ee
In addition, the gravitational backreaction of the scalar fluctuation reads
\be		\la{res:irscalariny}
A  \sim \fr{y}{R_{ir}} +  d_3 e^{- 2 \D'_- y /R_{ir}} \ls1 + \cdots \rs  + d_4 e^{- 2 \D'_+ y/R_{ir}}   \ls1 + \cdots \rs ,
\ee
where $d_3$ and $d_4$ are proportional to $c_3^2$ and $c_4^2$.

\begin{figure}
\centering
\subfigure[][]{\includegraphics[width=0.49\textwidth]{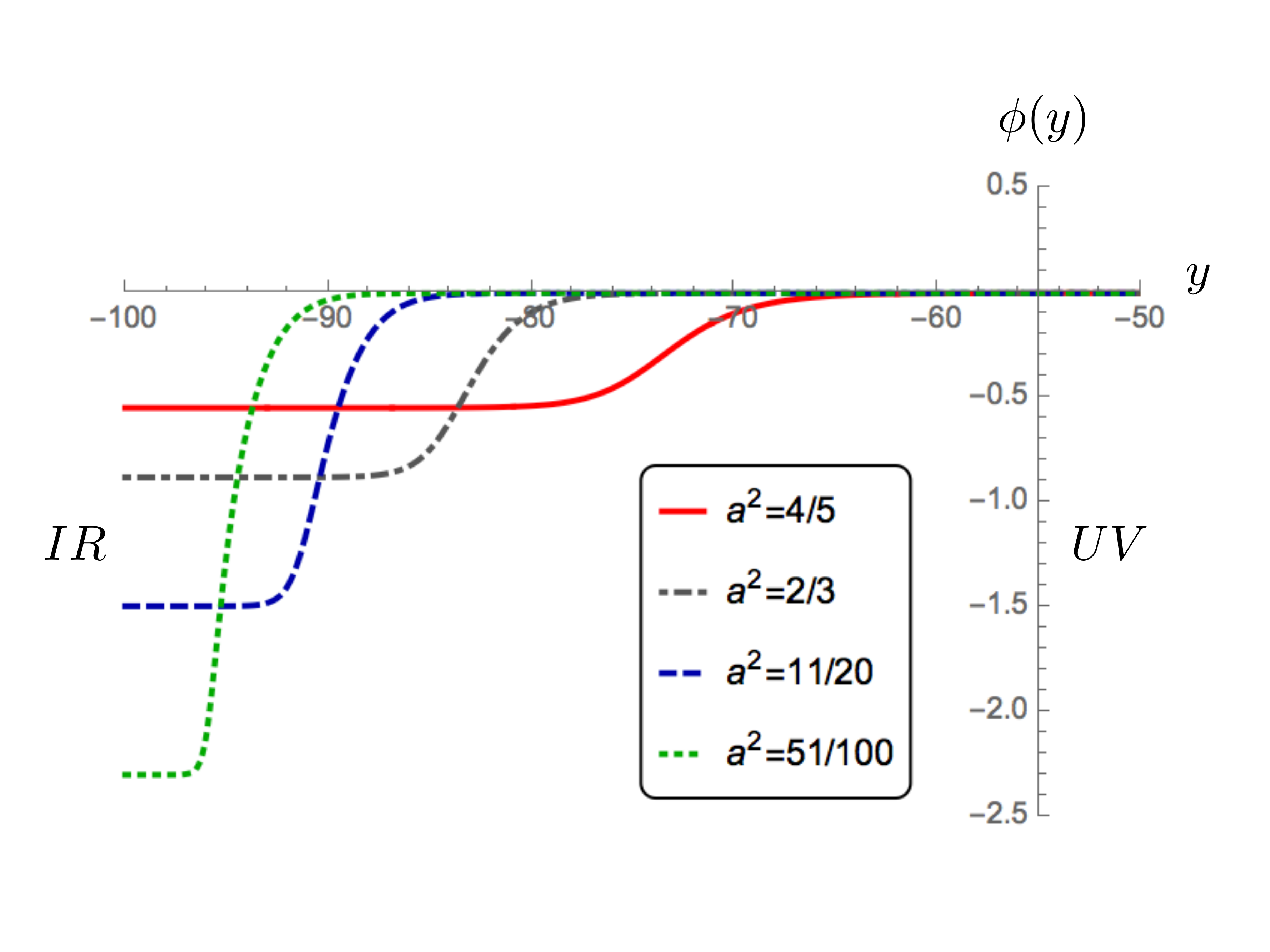}}
\subfigure[][]{\includegraphics[width=0.49\textwidth]{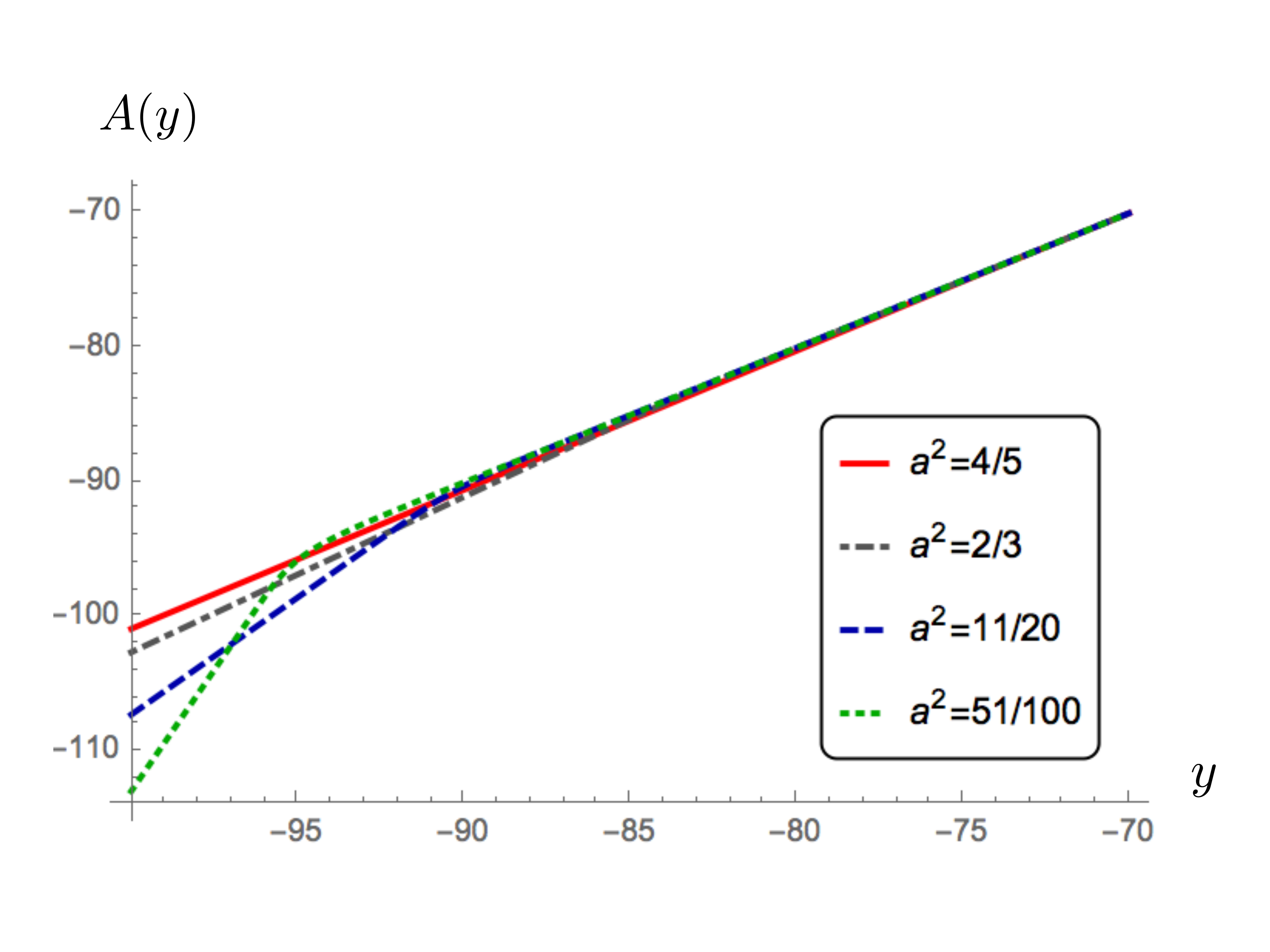}}
\caption{(Colour online) (a) The solutions $\phi$ connecting the two conformal fixed points for several different parameters $a$. (b) The domain wall behaviour of the metric function $A(y)$ for the different parameter $a$. }
\label{GeometricFlow}
\end{figure}

Above we showed that for $1/2 <a^2 <1$, the gravity we considered allows two different AdS geometries relying on the value of $\ph$. The one defined at $\ph=0$ is unstable and the other at the local minimum becomes stable. Can we find a new solution interpolating smoothly those stable and unstable AdS geometries? In Fig. 2, we show that there exists a geometric solution interpolating the local maximum and minimum smoothly. According to the AdS/CFT correspondence, such a geometry can be mapped to a deformation of a UV CFT which allows a new IR fixed point. More precisely, the unstable asymptotic AdS geometry with the AdS radius $R_{uv}$ is identified with the UV CFT because the UV fixed point becomes unstable under a relevant deformation. In this case, the relevant deformation operator leads to a nontrivial RG flow and makes the UV CFT become a new theory at low energy scale. If the deformation allows an IR fixed point, a new CFT appears at the IR fixed point which is again dual of another AdS space. Such an RG flow is holographically described by a geometry interpolating the two AdS geometries. The interpolating solutions in Fig.\ref{GeometricFlow} show such UV and IR features. The scalar field has $\ph=0$ at UV and $\ph=\ph_{ir}$ at IR with $\ph'=0$ and the corresponding metric $A(y)$ is given by a function linear to $y$ with the AdS radii, $R_{uv}$ and $R_{ir}$ at the UV and IR fixed points. In the intermediate scale, those two CFT features are smoothly connected, as expected.

If $a$ is given by a value in the range of $1/2 <a^2 <1$, we showed that there exists a geometric solution interpolating two different AdS geometries. At the fixed points of a nontrivial RG flow, there is an important quantity characterizing a CFT. That is called a central charge and represents the degrees of freedom of a CFT. According to the AdS/CFT correspondence, the central charge of a two-dimensional CFT is related to the quantities of the dual geometry
\be
c_{CFT}  = \fr{3 R_{AdS}}{2 G} ,
\ee
where $R_{AdS}$ means a AdS radius of a three-dimensional AdS space\cite{Brown:1986nw}. For the interpolating solution we found above, the relation of the central charge gives rise to 
\be		\la{res:UCcc}
c_{uv} = \fr{3 R_{uv}}{2 G}  ,
\ee
at the UV fixed point, while the central charge of the IR CFT becomes 
\be          \la{res:IRcc}
c_{ir} = \fr{3 R_{ir}}{2 G} .
\ee

In order to describe the degrees of freedom at the intermediate energy scale where the conformal symmetry is broken, we think of generalization of the central charge called a $c$-function. The $c$-function must be reduced to the previous central charges at the fixed points. Intriguingly, it has been argued that such a $c$-function monotonically decreases along the RG flow. This was called the $c$-theorem. On the dual gravity side, the nontrivial $c$-function of the dual field theory was realized in the holographic renormalization procedure \cite{Deger:2002hv,Gubser:2000nd}
\be
c=   \fr{3}{2 G} \fr{1}{ A'} ,
\ee
which reproduces and smoothly connects the above results at the fixed points, \eq{res:UCcc} and \eq{res:IRcc}. Since $R_{uv} > R_{ir}$ for the interpolating solution we found, the ratio of the central charges at the two fixed points is always smaller than $1$
\be
\fr{c_{ir}}{c_{uv}} = \fr{R_{ir}}{R_{uv}}  < 1 .
\ee
This indicates that the IR CFT always has a smaller central charge than that of the UV CFT, which is consistent with the $c$-theorem\cite{Zamolodchikov:1986gt,Barnes:2004jj}.

In order to prove the $c$-theorem, we further need to check whether the $c$-function monotonically decreases along the RG flow even at the intermediate energy scale. To do so, it is worth noting that we see from the recombination of \eq{eq:UVflucmetric} that $A''$ is always negative. Using this fact, the derivative of $c$ with respect to $y$ is always given by a positive value. This can be written in the following way with an additional minus sign
\be
- \fr{d c }{d y} =  \fr{3 A''}{2 G A'^2} = - \fr{3 \ph'^2}{2 G a^2 A'^2}  < 0 ,
\ee
where $-dy$ indicates the direction of the RG flow. This result proves that the proposed $c$-function in the dual gravity satisfies the $c$-theorem even at the intermediate energy scale.  Near the UV fixed point ($y \gg R_{uv} $), the $c$-function changes as 
\be
c = \fr{3 R_{uv}}{2 G} \ls 1 - \fr{(1-a^2) c_1^2}{a^2}  \ e^{ - 4 (1 - a^2) y/R_{uv}}  + \cdots \rs,
\ee
while in the IR regime  ($y \ll - R_{ir} $) it behaves as
\be
c = \fr{3 R_{ir}}{2 G} \ls 1 + 2 d_3 \D_-'  \ e^{ - 2  \D_-'  y/R_{ir}}  + \cdots \rs,
\ee
where $d_3 <0$ because $\D_-'  <0$ and $d c/dy > 0$.

%%%%%%%%%%%%
\section{Holographic entanglement entropy}
\label{sec:4}

In the previous section, we have studied the holographic RG flow of a UV CFT deformed by a relevant scalar operator which is dual to the geometry interpolating two different AdS spaces. In this section, we investigate how the entanglement entropy deforms along the RG flow. The RG flow of the entanglement entropy is related to the real space RG flow. Except several specific cases, in general, it is not easy to calculate the entanglement entropy of an interacting quantum field theory. However, the AdS/CFT correspondence can provide a new technique to look into the entanglement entropy of an interacting theory and its RG flow. In this section, we will investigate the RG flow of the entanglement entropy on the  previous interpolating geometry holographically.   

\subsection{Entanglement entropy and $c$-function near the UV fixed point}

According to the Ryu-Takayangi formula, the entanglement entropy of the dual field theory can be evaluated by calculating the area of the minimal surface whose boundary is coincident with the entangling surface distinguishing two subsystems of the boundary field theory. For the dual field theory of the previous interpolating geometry, the entangling surface is given by two points which divide the space of the dual field theory into two systems. Now, we assume that the entangling points are located at $x=\pm l/2$. In this case, the dual field theory is divided into a subsystem defined at $-l/2 \le x \le l/2$ and its complement. Then, the entanglement entropy is represented by the minimal surface extended to the dual interpolating geometry with connecting two entangling points. From \eq{metric:adsy}, the corresponding entanglement entropy is governed by
\be		\la{eq:orgHEE}
S_E = \fr{1}{4 G} \int_{-l/2}^{l/2}  dx \sqrt{e^{2 A(y)} + y'^2} ,
\ee 
where $y$ is given by a function of $x$. Denoting the turning point of the minimal surface as $y_*$ at which $y'$ vanishes, the conserved quantity of the above entanglement entropy allows us to represent the subsystem size and entanglement entropy as functions of the turning point. After some calculations, the size of the subsystem can be expressed in terms of the turning point
\be			\la{eq1:subsystemsize}
l =2 \int^\infty_{y_\ast} dy \ \frac{e^{A_*}}{e^{ A }\sqrt{e^{2 A}-e^{2 A_\ast }}},
\ee
where $A_\ast$ indicates the value of A at $y=y_*$. In addition, the entanglement entropy is determined by the following integral relation
\be	 \label{srt}
S_{E}=\frac{1}{2 G}\int_{y_\ast}^{y_{uv}} dy \ \frac{e^{ A }}{\sqrt{e^{2 A }-e^{2 A_* }}}.
\ee
Here we introduced an appropriate UV cutoff $y_{uv}$ because the entanglement entropy diverges as $y_{uv} \to \infty$.

Let us first take into account a small subsystem size which describes the UV entanglement entropy. In this UV limit, the turning point has a large value satisfying $y_*/R_{uv} \gg 1$ and the geometric solution except for $a^2=1/2$ can be well approximated by
\be
A = \fr{y}{R_{uv}}   - \fr{  c_1^2}{4 a^2} \  e^{ - 4 (1 - a^2) y/R_{uv}}  \ls1 + \cdots \rs   -\frac{c_2^2 }{4 a^2 }  \ e^{- 4 a^2 y/R_{uv}}  \ls1 + \cdots \rs .
\ee  
Substituting this perturbative geometric solution into \eq{eq1:subsystemsize} and \eq{srt} and performing the integrals, we can finally obtain the perturbative entanglement entropy in terms of the subsystems size
\be
S_{E}=S_{E}^{(0)}+  S_{E}^{(1)}+\cdots  .
%=\frac{1}{4G}\biggl( {\cal{A}}_{{Min}}^{(0)}+\varepsilon^2{\cal{A}}_{{Min}}^{(2)}+\cdots\biggr),
\ee
Here $S_{E}^{(0)}$ indicates the leading contribution from the AdS space with the AdS radius $R_{uv}$, while $S_{E}^{(1)}$ means the first correction from the scalar deformation. The leading contribution reads
\be
S_E^{(0)}%=\frac{1}{2G}\log\frac{l}{\epsilon_{uv}}
=\frac{c_{uv}}{3}\log\frac{l}{\epsilon_{uv}},
\ee
with
\be
c_{uv} = \fr{ 3 R_{uv}}{2 G} \quad {\rm and} \quad \epsilon_{uv} =  R_{uv} e^{- y_{uv}/R_{uv}}  .
\ee
In this case, the central charge of the dual UV CFT is derived by varying the entanglement entropy with respect $\log l$ \cite{Myers:2012ed}
\be
c_{uv} = \fr{3 \ d S_E^{(0)}}{d \log l}  .
\ee
Note that this leading contribution is independent of the value of $a$ which characterizes the scalar deformation.

Now, let us consider the first correction caused by the scalar deformation. Relying on the value of $a$, the corresponding contribution to the entanglement entropy can show different behaviors. For $1/2 <a^2 < 1$, $c_1$ corresponding to the source becomes dominant. Thus, the first correction to the entanglement entropy leads to
\be		\la{res:c1}
%y_\ast&=&\log\frac{2}{l}+c_1^2\,l^{4(1-a^2)}\,\frac{\Gamma(7/2-2a^2)-2(1-a^2)\sqrt{\pi} \Gamma(3-2a^2)}{2^{2(3-2a^2)}a^2\,\Gamma(7/2-2a^2)},
%\nonumber \\
S_{E}^{(1)} = - \frac{ \lc \sqrt{\pi } \Gamma \left(3-2 a^2\right) - 2 \Gamma \left(\frac{7}{2}-2 a^2\right)\rc  R_{{uv}} c_1^2}{ 2^{4(2-a^2)} a^2 G  \Gamma \left(\frac{7}{2}-2 a^2\right)}  \fr{l^{4(1-a^2)}} {R_{{uv}}^{4(1-a^2)}} + \cdots  .
% -  \fr{1}{G} \frac{\sqrt{\pi}\,\Gamma(3-2a^2) c_1^2 }{2^{4(1-a^2)}a^2\,\Gamma(7/2-2a^2)} \,l^{4(1-a^2)}.
\ee
Since $S_E^{(1)}$ is negative for $1/2 <a^2 < 1$, this result implies that the entanglement entropy in the UV limit monotonically decreases along the RG flow. More precisely, the UV central charge depending on the small subsystem size is given by
\be
c = c_{uv} \ls 1 - \frac{ \left(1-a^2 \right)   \lc \sqrt{\pi } \Gamma \left(3-2 a^2\right)-2 \Gamma \left(\frac{7}{2}-2 a^2\right)\rc c_1^2   }{2^{5-4 a^2} a^2 \Gamma \left(\frac{7}{2}-2 a^2\right)} \fr{ l^{4(1-a^2)}}{R_{{uv}}^{4(1- a^2)}} +\cdots \rs .
%\fr{3 \ d S_E}{d \log l} = c_{{uv}} \ls 1- \frac{\sqrt{\pi }    c_1^2}{ 2^{4(1- a^2)} } \fr{\left(1 - a^2 \right) \Gamma \left(3-2 a^2\right)  }{a^2 \Gamma \left(\frac{5}{2}-2 a^2\right)}  \fr{l^{4 (1- a^2)  } }{R_{uv}^{4(1- a^2)}}  + \cdots \rs .
\ee
If $dc/dl = 0$ at the UV fixed point, the $c$-function is called stationary \cite{Zamolodchikov:1986gt,Liu:2012eea,Kim:2014yca,Kim:2014qpa,Park:2015afa,Park:2015dia,Kim:2018mgz,Kim:2015rvu}. Above, the resulting $c$-function is stationary for $1/2 <a^2  < 3/4$, while it is not for $3/4 \le a^2 <1$.

In the other range of $a^2$, similarly, we can summarize the UV entanglement entropy and its central charge as follows.
\begin{itemize}
\item  For $a^2 >1$, as mentioned before, the vacuum solution allows a stable AdS geometry. In this case, the stability of the AdS geometry implies that $c_1$ vanishes. If not, the graviational backreacton of the $c_1$ term can modify the background AdS geometry because its asymptotic value diverges exponentially. Despite this fact, if we take a nonvanishing $c_1$, its contribution  gives rise to the entanglement entropy proportional to $S_{E}^{(1)} \sim 1/l^{4 (a^2-1) }$.
For $a^2 > 1$ the first correction in the UV limit leads to a power-law divergence which is more severe than the logarithmic divergence of a two-dimensional CFT.  As a consequence, $c_1$ must be taken to be zero for $a^2 >1$ in order to obtain a stable AdS geometry. Anyway, the $c_2$ term is rapidly suppressed at the boundary, so that this term with a vanishing source term ($c_1=0$) does not ruin the background AdS geometry. Its contribution to the entanglement entropy becomes
\be			\label{res:eec2}
%y_\ast&=&\log\frac{2}{l}+c_2^2\,l^{4a^2}\,\frac{\Gamma(3/2+2a^2)-4a^4\sqrt{\pi}\,\Gamma(2a^2)}{2^{2(1+2a^2)}a^2\,\Gamma(3/2+2a^2)},
%\nonumber
%\\
S_{E}^{(1)}  = - \frac{  \lc  2 \Gamma \left(2 a^2+\frac{3}{2}\right) - \sqrt{\pi } \Gamma \left(2 a^2+1\right) \rc R_{{uv}} c_2^2 }{ 2^{4(1+ a^2)} a^2 G   \Gamma \left(2 a^2+\frac{3}{2}\right)} \fr{l^{4 a^2}}{R_{{uv}}^{4 a^2}} + \cdots ,
%-\frac{\sqrt{\pi }   \Gamma \left(2 a^2\right) c_2^2 R_{{uv}}}{2^{4 a^2+2}  \Gamma \left(2  a^2+\frac{1}{2}\right) G }  \fr{l^{4 a^2} }{R_{{uv}}^{4 a^2}},
\ee
which for $a^2 > 1$ is rapidly suppressed in the UV limit. The corresponding $c$-function reduces to
\be  		\label{res:cc2}
c =  c_{{uv}} \left( 1-  \frac{ \lc 2 \Gamma \left(2 a^2+\frac{3}{2}\right) - \sqrt{\pi } \Gamma \left(2 a^2+1\right) \rc  c_2^2  }{2^{1+4 a^2} \Gamma \left(2 a^2+\frac{3}{2}\right)} \fr{l^{4 a^2} }{R_{{uv}}^{4 a^2}} + \cdots \right) .
%c_{{uv}} \left(1-\frac{\sqrt{\pi }    \Gamma \left(2 a^2+1\right)   c_2^2 }{ 2^{4 a^2} \Gamma \left(2 a^2+\frac{1}{2}\right)} \fr{l^{4 a^2}}{R_{uv}^{4 a^2}} + \cdots \right) .
\ee
This result shows that the $c$-function is stationary for $a^2 > 1$ because of the absence of the contribution from the source term, $c_1$.

%\tcb{Because of the absence of the leading contribution from $c_1$, the UV fixed point is always stationary.} \tcr{Is this deformation allowed? It would be modified the IR geometry, so the resulting IR geometry is not an AdS space. This is inconsistent with the stable vacuum. How about the entanglement entropy? }

\item For $a^2=1$, the $c_1$ term becomes a constant and $c_2$ corresponds to the vev of a marginal operator. In this case, the constant $c_1$ term does not give any contribution to the entanglement entropy. As a result, the marginal scalar deformation yields
\be
S_{E}^{(1)} = -\frac{7 R_{{uv}} c_2^2}{1920 G }  \fr{ l^4}{R_{{uv}}^4} + \cdots,
\ee
and the $c$-function reads
\be
c = c_{{uv}} \ls 1 -\frac{7 c_2^2 }{240 } \fr{ l^4}{R_{{uv}}^4}  + \cdots \rs,
\ee 
which is stationary.

\item For $1/2 <a^2 < 1$, as explained before, the $c$-function is stationary for $1/2 <a^2  < 3/4$ but not for $3/4 \le a^2 <1$.

\item  When $a^2 = 1/2$, the role of $\bar{c}_1$ and $\bar{c}_2$ in \eq{res:a2half} becomes ambiguous because they are degenerate. In this case, the first correction comes from the $\bar{c}_2$ term. After some calculation, the resulting entanglement entropy deformed by a double trace operator reads
\be 
S_E = \frac{R_{{uv}} }{2  G}  \log  \fr{l}{\epsilon _{{uv}}}   -  \frac{\lc  36   \log ^2 \ls l/ R_{{uv}} \rs
- 60 \log \ls l/ R_{{uv}} \rs +56 -3 \pi ^2\rc R_{{uv}} c_2^2   }{864 G }  \fr{l^2}{R_{{uv}}^2} + \cdots ,
\ee
and the corresponding $c$-function becomes
\be
c&=& c_{{uv}} \ls 1  -  \frac{  \lc 36 \log^2  \ls l /R_{uv} \rs  - 24 \log \ls l /R_{uv} \rs +26 -3 \pi ^2 \rc c_2^2 }{216 }  \fr{ l^2}{R_{{uv}}^2}   + \cdots \rs  .
\ee
This results shows that the UV fixed point for $a^2=1/2$ is not stationary.

\item For $0 <a^2 < 1/2$, $c_1$ and $c_2$ exchange their role and the $c_2$ term becomes dominant. In this case, the first correction to the entanglement entropy and $c$-function are given by \eq{res:eec2} and \eq{res:cc2}. From \eq{res:cc2}, we can see that the $c$-function is stationary at the UV fixed point for $0 < a^2 < 1/4$ but not for $1/4 \le a^2 < 1/2$. 

\end{itemize}

\subsection{Central charge of the dual IR CFT}

Now, let us discuss the entanglement entropy in the IR regime. For a pure AdS$_3$ in \eq{metric:adsy} with $A(y)=y/R$ which is dual to a two-dimensional CFT,  it has been well known that the entanglement entropy of a two-dimensional CFT can exactly be reproduced by the holographic calculation of the minimal surface area\cite{Ryu:2006bv}. From the dual field theory point of view, the turning point of the minimal surface corresponds to the lowest energy scale we can observe by using the holography. Therefore, if we want to figure out the IR entanglement entropy and its RG flow, we take a very small turning point satisfying $y_*  \ll - R_{ir}$ where the IR geometry is characterized by $R_{ir}$. This is equivalent to take a large subsystem with $l \gg R_{ir}$. In general, the holographic entanglement entropy is a non-local quantity. On the dual geometry side, this is because the minimal surface is extended from the IR energy described by $y_*$ to the UV energy denoted by $y_{uv}$. Due to this fact, it is not easy to evaluate an exact and analytic result of the IR entanglement entropy. In this section, we derive the approximated IR entanglement entropy near the IR fixed point. Because of the restoration of the conformal symmetry near the IR fixed point, we can expect that the known entanglement entropy of CFT again appears with a different central charge representing the IR degrees of freedom.
 
As mentioned before, let us assume that the turning point is located at the IR fixed point with $y_*  \ll - R_{ir}$. Then, $e^{A(y_*)}$ can be well approximated by $e^{y_*/R_{ir}}$ in the IR regime. Now, we divide the integral range of $l$ in \eq{eq1:subsystemsize} and $S_E$ in \eq{srt} into two parts
\be			
l &=& 2 \ls \int_{y_\ast+ \e }^{y_{uv}} dy  + \int_{y_\ast}^{y_\ast+ \e} dy \rs  \ \frac{ e^{y_*/R_{ir}}}{ e^{2 A }  \sqrt{1-e^{2 ( {y_*/R_{ir}} - A )}}}, \nn\\
S_{E}&=&\frac{1}{2 G} \ls \int_{y_\ast + \e}^{y_{uv}} dy +\int_{y_\ast}^{y_\ast+ \e}  dy \rs \ \frac{1}{\sqrt{1-e^{2 ( {y_*/R_{ir}} - A  )}}} ,   
\ee
Assuming that $\e$ is in the range of $R_{ir} \ll \e \ll |y_*|$, the inside of the square root above can be approximated by $1$ for $y_* + \e \le  y \le y_{uv}$ because $A(y) - y_*/R_{ir} \gg 1$ in this region. On the other hand, $A(y)$ for $y_* \le  y \le y_* + \e \ll - R_{ir}$ is approximated by $A(y) \approx y/R_{ir}$. Using these facts, the subsystem size becomes approximately
\be
l \approx 2 \int_{y_\ast+ \e }^{y_{uv}} dy  \ e^{y_*/R_{ir} - 2 A(y)}
+2  \int_{y_\ast}^{y_\ast+ \e} dy    \ \frac{ e^{y_*/R_{ir}}}{e^{ 2 y/R_{ir}}  \sqrt{1-e^{- 2 ( y - {y_*})/R_{ir}} }} .
\ee
In this case, most of the contribution to the subsystem size comes from the second integral because the integrand of the first integral is suppressed exponentially ($e^{y_*/R_{ir} - 2 A(y)} \ll 1$) for $y_* + \e \le  y \le y_{uv}$. The integration of the second term leads to
\be
l \approx R_{ir} e^{ - y_*/R_{ir}} ,
\ee
where $\e/R_{ir} \gg 1$ was used. Similarly, the entanglement entropy can also be rewritten as
\be			
S_{E} \approx  \frac{1}{2 G}  \int_{y_\ast + \e}^{y_{uv}} dy + \frac{1}{2 G} \int_{y_\ast}^{y_\ast+ \e}  dy  \ \frac{1}{\sqrt{1-e^{- 2 ( y - {y_*})/R_{ir}} }} .   
\ee
Performing the integral and rewriting the result in terms of $l$, we finally obtain  
\be
S_{E} \approx \frac{y_{uv}}{2 G}   - \frac{y_*}{2 G}  = \frac{y_{uv}}{2 G}  +  \frac{R_{ir}}{2 G} \log \fr{l}{R_{ir}} ,
\ee 
where the first term indicates the UV divergent term. The second term corresponding to the leading IR entanglement entropy shows that the IR entanglement entropy near the IR fixed point has the exact same form as the one derived from the UV fixed point. This is because of the restoration of the conformal symmetry at the IR fixed point. Furthermore, the degrees of freedom of the IR CFT can be represented by the IR central charge given by
\be			\la{res:ircentralcharge}
c_{ir} =\fr{3 d S_{E}}{d \log l} =  \fr{3 R_{ir}}{2 G} ,
\ee
which is exactly the expected form.

\section{RG flow of the entanglement entropy}
\label{sec:5}

In the previous section, we have studied the entanglement entropy at the UV and IR fixed points and showed that the central charges derived from the holographic entanglement entropy at the two fixed points lead to the consistent results expected from the dual CFT. In this section, we will further investigate how the central charge evolves at the intermediate energy scale along the RG flow analytically and numerically.

\subsection{Thin-wall approximation}

In the model we considered, the scalar field has a kink-type profile as shown in Fig. 2(a). Thus, we can apply the thin-wall approximation in order to understand analytically the RG flow of the central charge. In \cite{Albash:2011nq,Myers:2012ed}, the similar thin-wall approximation has been taken into account. It has been argued that there can exist a region where the $c$-theorem may be broken. Due to this fact, a phase transition occurs and connects the two UV and IR CFTs without breaking of the $c$-theorem \cite{ Myers:2012ed}. In this section, we apply the similar thin-wall approximation with a different prescription from the one used in \cite{ Myers:2012ed}. We impose a junction condition requiring that the minimal surface extended to the dual geometry must be continuous. The thin-wall approximation with this condition looks more natural and shows that there is no phase transition. In other words, the $c$-theorem is not broken in the entire range of the RG flow. In order to check whether the thin-wall approximation with the junction condition is valid, we further calculate the exact RG flow of the central charge numerically and then compare those two results in the next section.

\subsubsection{In the UV region with a small subsystem size}

In order to apply the thin-wall approximation, let us first consider the junction of two AdS spaces with different AdS radii, $R_{uv}$ and $R_{ir}$, which is a simple approximation of the previous interpolating geometry. Then, two AdS metrics can be represented as
\be			\la{ans:UVIRgeometries}
A(y) &=&  \fr{y}{R_{uv}} \quad {\rm for} \ y > y_w  , \nn\\
&=& \fr{y-y_w}{R_{ir}}+\fr{y_w}{R_{uv}}  \quad {\rm for} \ y < y_w  ,
\ee
where $y_w$ indicates the position of the thin-wall. Notice that the first term in the second IR metric factor was introduced for continuity of the metric at the wall.

We first consider the case with a small subsystem size in which the corresponding minimal surface does not touch the wall. In other words, the turning point of the minimal surface is always larger than the wall's position, $y_* > y_w$. Since the holographic entanglement entropy cannot measure IR physics below $y_*$, it is sufficient to take into account only the region with $y_* < y < \infty$ where the dual geometry is the AdS space with the AdS radius $R_{uv}$. Physically, this clarifies the UV feature of the entanglement entropy. Denoting the trajectory of the minimal surface by $\lc \bar{x}, \bar{y}\rc$, it can be determined by solving the following integral equation derived from \eq{eq1:subsystemsize} 
\be		
\int_{\bar{x}}^{l/2} dx &=& \int^\infty_{\bar{y}} dy \ \frac{e^{y_*/R_{uv}}}{e^{ y/R_{uv}}\sqrt{e^{2 y/R_{uv}}-e^{2 y_*/R_{uv} }}}.
\ee
In particular, the turning point corresponds to $\lc \bar{x}, \bar{y}\rc = \lc 0, y_*\rc$ and $y_*$ can be expressed in terms of the subsystem size, $y_* = R_{uv} \log \ls 2 R_{uv}/l\rs$. Performing this integral exactly, $\bar{y}$ can be determined as a function of $\bar{x}$ 
\be	 	\la{eq:trajyitox}
\bar{y} = \frac{R_{uv}}{2}  \log \fr{4 R_{uv}^2}{l^2 - 4 \bar{x}^2} ,
\ee
which reproduces the turning point at $\lc \bar{x}, \bar{y}\rc = \lc 0, y_*\rc$. Rewriting $\bar{x}$ in terms of $\bar{y}$ leads to
\be		\la{eq:trajxitoy}
\bar{x} = \fr{1}{2} \sqrt{l^2 - 4  R_{uv}^2 e^{-2 \bar{y}/R_{uv}}} .
\ee
After substituting the trajectory in \eq{eq:trajyitox} into the entanglement entropy formula in \eq{eq:orgHEE}, performing the integral gives rise to the following entanglement entropy
\be 
S_{E} = \fr{y_{uv}}{2 G}  +  \fr{R_{uv}}{2 G}  \log \fr{l}{R_{uv}}   ,
\ee
where $ y_{uv} $ means the UV cutoff in the $y$-coordinate. This UV result is consistent with the previous result of the UV CFT, as it should do.

\subsubsection{In the IR limit with a large subsystem size}

Now, let us take into account the IR entanglement entropy. When the subsystem size is sufficiently large, the minimal surface is extended to the IR regime ($y_* < y_w$) as well as the UV regime ($y_* > y_w$). In the UV region, the minimal surface's trajectory is described by \eq{eq:trajyitox} or \eq{eq:trajxitoy}. Since the UV part covers only the range of  $y_w < y_* < \infty $, the minimal surface meets the thin-wall at $\lc x_w, y_w\rc$ with
\be		\la{re:trajinUV}
x_w = \fr{1}{2} \sqrt{l^2 - 4  R_{uv}^2 e^{-2 y_w/R_{uv}}} .
\ee

In the IR region, on the other hand, the trajectory of the minimal surface is determined by
\be        \la{re:trajinIR}
\bar{x}=\fr{1}{2}\sqrt{ l_w^2-4 R_{ir}^2 e^{-2\bar{y}/R_{ir}} e^{2y_w(R_{uv}-R_{ir})/R_{uv}R_{ir}}},
\ee
where $\bar{y}$ lies in the range of $y_* < y <y_w$. Above $l_w$ indicates the subsystem size on the wall located at $y=y_w$. In order to obtain a continuous trajectory of the minimal surface, we need to require two constraints. First, the turning point is located at $\lc \bar{x}, \bar{y}\rc = \lc 0, y_*\rc$ in \eq{re:trajinIR}. This fixes the undetermined $l_w$ in terms of $y_*$
\be  \la{res:barl}
l_w=2 R_{ir} e^{-\bar{y}_*/R_{ir}} e^{y_w(R_{uv}-R_{ir})/R_{uv}R_{ir}}.
\ee
Second, the continuity of the minimal surface on the wall requires that both \eq{re:trajinUV} and \eq{re:trajinIR} lead to the same value at $  \bar{y} = y_w $. This requirement together with \eq{res:barl} gives rise to
\be
l=2 \sqrt{ R_{uv}^2 e^{-2 y_w /R_{uv}} + R_{ir}^2 e^{-2y_*/R_{ir}} e^{2y_w(R_{uv}-R_{ir})/R_{uv}R_{ir}} -R_{ir}^2 e^{-2y_w/R_{uv}} } .
\ee
This relation shows that the subsystem size can be determined in terms of $y_*$ and $y_w$. When $l$ and $y_w$ are given, in other words, the position of the turning point is given by
\be
y_* = \fr{R_{uv}-R_{ir}}{R_{uv}}y_w-\fr{R_{ir}}{2} \log \ls \fr{l^2}{4 R_{ir}^2}  + e^{- 2 y_w/R_{uv}} - \fr{R_{uv}^2}{R_{ir}^2}  e^{-2 y_w/R_{uv}} \rs .
\ee

The entanglement entropy described by the trajectory of the minimal surface is represented as
\be			
S_{E} = \frac{1}{2 G}  \int_{y_w}^{\infty} dy \ \frac{1}{\sqrt{1-e^{- 2 ( y - {y_{**}})/R_{uv}} }}+ \frac{1}{2 G} \int_{y_\ast}^{y_w}  dy  \ \frac{1}{\sqrt{1-e^{- 2 ( y - {y_*})/R_{ir}} }} , 
\ee
where $y_{**}$ corresponds to the turning point when the minimal surface extends only to the UV geometry with $R_{uv}$
\be
y_{**} = R_{uv} \log \fr{2 R_{uv}}{l} .
\ee
Then, the resulting entanglement entropy after performing the integral reads
\be
S_{E} &=& \fr{1}{2G} \lb \Lambda+R_{uv} \log 2-y_\ast -R_{uv}\log \left(1+ \sqrt{1-e^{2   \left(y_{**}-y_w\right)/R_{uv}}} \right)     \rp \nn\\
&& \qquad \lp +R_{ir} \log \left(1+ \sqrt{1-e^{2 \left(y_{*}-y_w\right)/R_{ir}}}\right) \rb
\ee
In the large $l$ limit, this entanglement entropy is expanded into
\bea \nonumber
S_{E} &=& \fr{1}{2G} \biggl [ \Lambda +R_{ir} \log l -\fr{R_{uv}-R_{ir}}{R_{uv}}y_w-R_{ir} \log R_{ir}  
\\ \nonumber
&+&\biggl(\fr{R_{uv}^3 e^{-2y_w/R_{uv}}-R_{ir}^3 e^{-2(2R_{uv}-R_{ir})y_w/R_{uv}R_{ir}}-2R_{ir}(R_{uv}^2-R_{ir}^2)e^{-y_w/R_{uv}}}{l^2} \biggr)
\biggr]
\\
&+& {\cal O} \ls \fr{1}{l^{3}} \rs.
\eea
From this result, we can see that the $c$-function near the IR fixed point behaves like
\be
c &=& c_{ir} \ls 1+ \frac{2R_{ir}(R_{uv}^2-R_{ir}^2)e^{-y_w/R_{uv}}-R_{uv}^3 e^{-2y_w/R_{uv}}+R_{ir}^3 e^{-2(2R_{uv}-R_{ir})y_w/R_{uv}R_{ir}}}{ R_{ir} } \fr{2}{l^2} \rs \nn\\
&&+ {\cal O} \ls \fr{1}{l^{3}} \rs ,
\ee
which reproduces the IR central charge \eq{res:ircentralcharge} in the limit of $l \to \infty$.

\subsection{Comparison with the exact numerical result}

\begin{figure}
\centering
\subfigure[][]{\includegraphics[angle=0,width=0.49\textwidth]{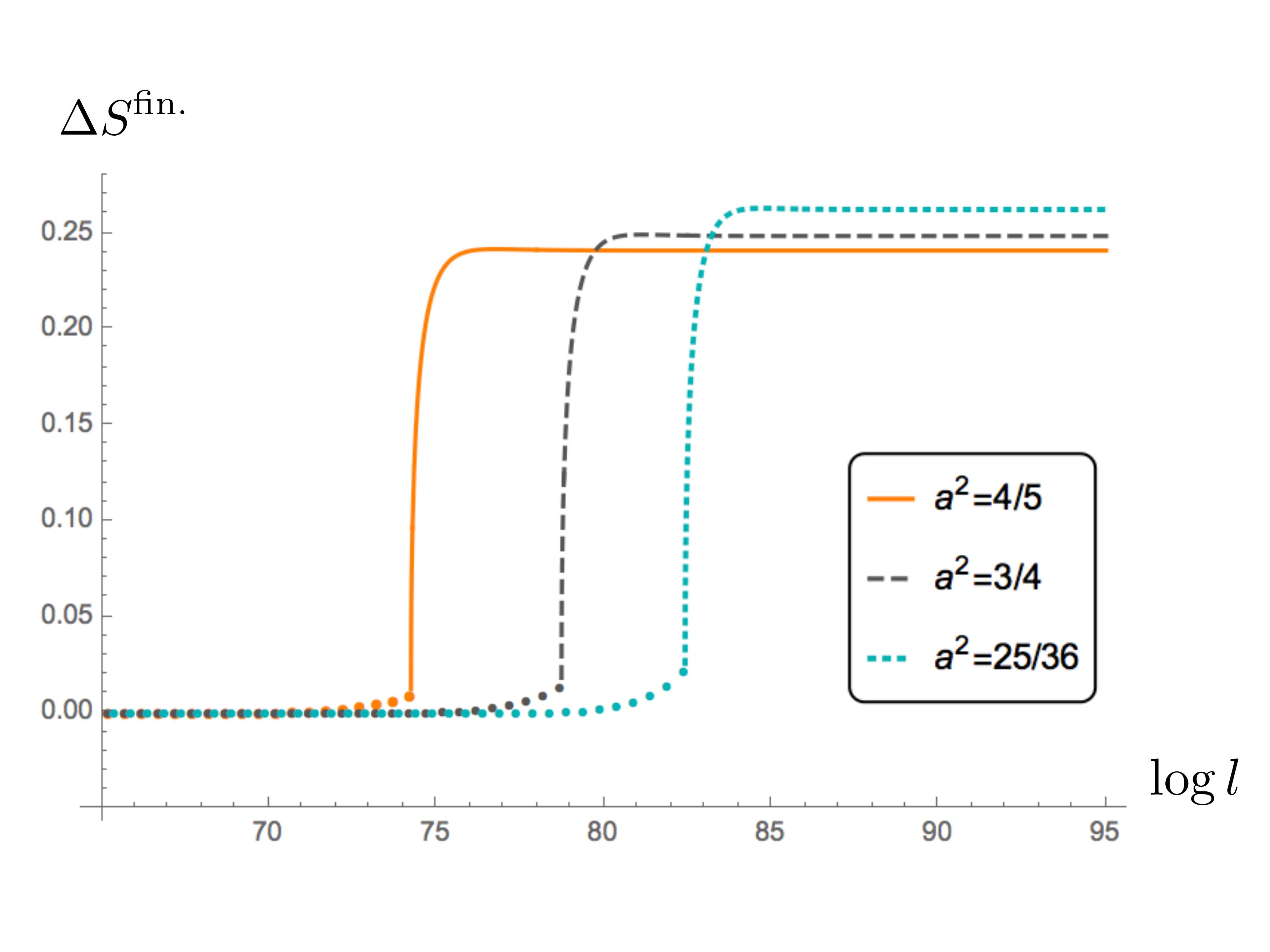}}
\subfigure[][]{\includegraphics[angle=0,width=0.49\textwidth]{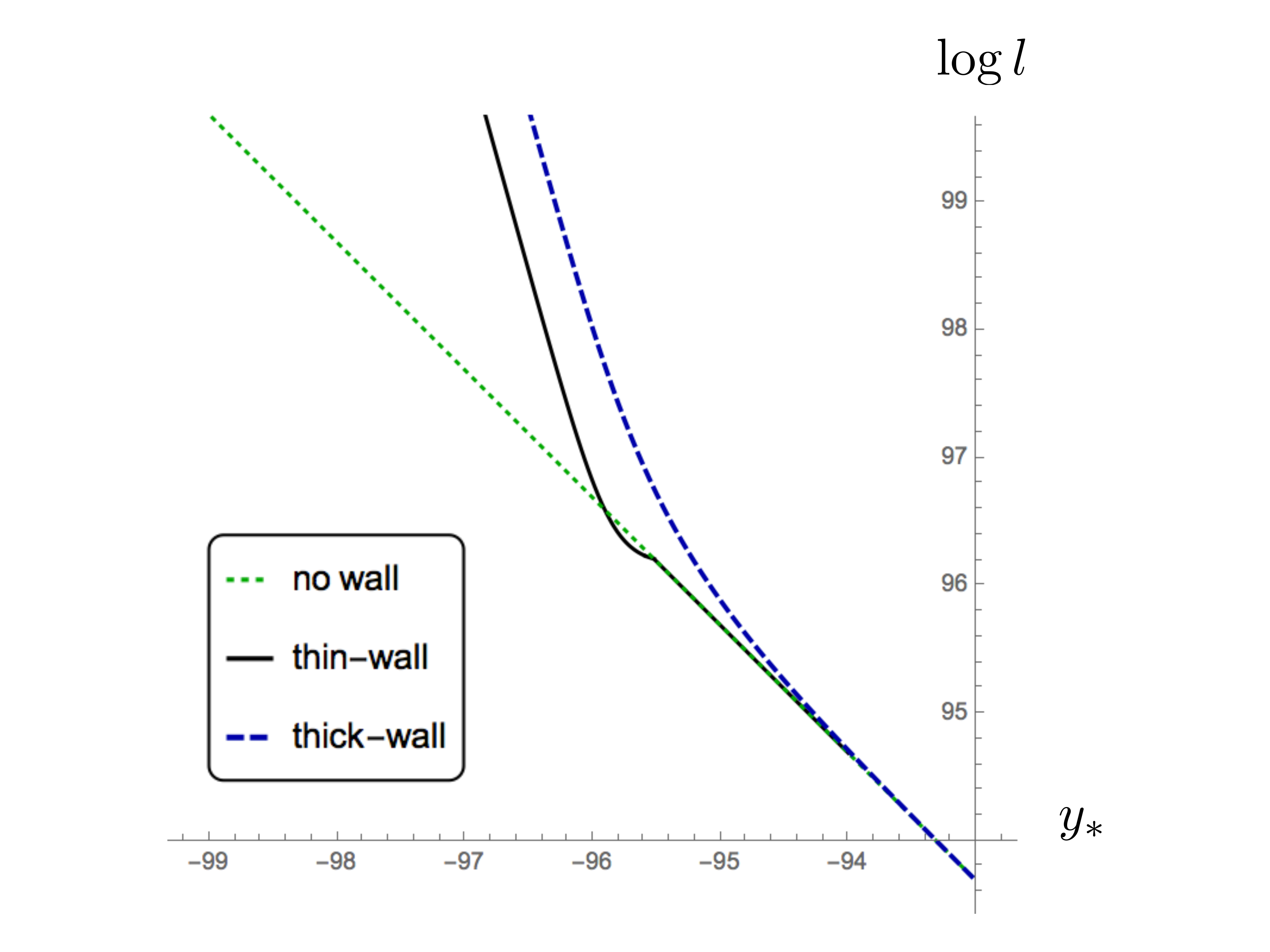}}
\caption{(color online). (a)  This plot illustrates different values between thin and thick-wall domain solutions, $\Delta S^{\text{fin.}}=S^{thin}-S^{thick}$. We present two regions separated by a position of thick-wall. For the ranges, which denoted by circle lines, $y < y_{\text{DW}}$, the thin-wall entropy are the same as the entanglement entropy of the pure AdS : $S^{thin}=S^{\text{AdS}}$. For $y > y_{\text{DW}}$, $\Delta S$ converges to specific finite values for large $l\gg 1$. (b) This plot illustrates the position of turning point for each different domain-wall solutions with $a^2=51/100$.}
\label{thickVSthin}
\end{figure}

Now, let us take into account an exact numerical $c$-function and compare it with the previous thin-wall approximation. The exact $c$-function for a deformed $AdS_3$ space can be formally written as  \cite{Myers:2012ed}
\be
 c =  \frac{3 dS_E}{d \log l}=\frac{3}{4G}\frac{l}{\gamma(l)} ,
\ee
where $\gamma(l)$ indicates a constant of motion relying on the subsystem size $l$. The constant of motion is generally given by a function of the turning point $y_*$ 
\be
\g (y_*) = e^{-  A_*}  .
\ee
However, the integral relation in \eq{eq1:subsystemsize}, in principle, allows us to rewrite $y_*$ as a function of $l$. Above, $\gamma(l)$ means the constant of motion expressed in terms of the subsystem size $l$ instead of $y_*$. In Ref. \cite{Myers:2012ed}, the similar thin-wall approximation with imposing the reality of the entanglement entropy has been taken into account. Under this prescription, the configuration of the minimal surface becomes folded and allows a region where the $c$-theorem is broken down. Because of this, it was argued that there exists a phase transition when the thickness of the wall is sufficiently thin. In the previous section, we applied the thin-wall approximation with a different prescription which looks more natural. Instead of the reality of the entanglement entropy, we imposed a natural junction condition, the continuity of the minimal surface on the wall, which guarantees the reality of the entanglement entropy. With the natural junction condition, we depict the entanglement entropy difference between the exact result and the thin-wall approximation in Fig. \ref{thickVSthin}(a). The result shows that the difference suddenly becomes large at the intermediate energy scale corresponding to the position of the wall, and that the difference remains a constant in the IR limit. This is because the background geometries used in the thin-wall approximation do not contain proper information about the deformation of the background geometry. Despite this error, the configuration of the minimal surface in Fig. \ref{thickVSthin}(b) shows that the position of the turning point monotonically decreases when the subsystem size increases. In other words, the minimal surface is not folded as the exact result shows in Fig. \ref{thickVSthin}(b).  The present thin-wall approximation with the natural junction condition shows the consistent result with the exact one in that there is no folding of the minimal surface. Therefore, there is no phase transition in the dual field theory of the interpolating geometry.

In Fig. 4(a), we investigate the change of the $c$-functions with the thin-wall approximation and exact numerical calculation. The results show that the thin-wall approximation studied before gives rise to the consistent result with the exact numerical result except the near of the wall. In Fig. \ref{thickVSthin}(a), we showed that the entanglement entropy derived from the thin-wall approximation has a constant difference from that of the exact numerical calculation in the IR limit. In this case, the constant difference implies the different which is independent fo the subsystem size. In spite of this constant difference, the $c$-functions derived by two different methods are well matched because the constant difference does not give any effect on the $c$-function. Therefore, the thin-wall approximation with the natural junction condition is a good approximation representing the correct $c$-function only with some error near the wall. Anyway, if we take $a^2 \approx 1$, the deformation of the background geometry becomes negligible. Thus, the thin-wall approximation is well matched to the exact numerical result (see Fig. 4(b)).

\begin{figure}
\centering
\subfigure[For $a^2=0.51$]{\includegraphics[angle=0,width=0.49\textwidth]{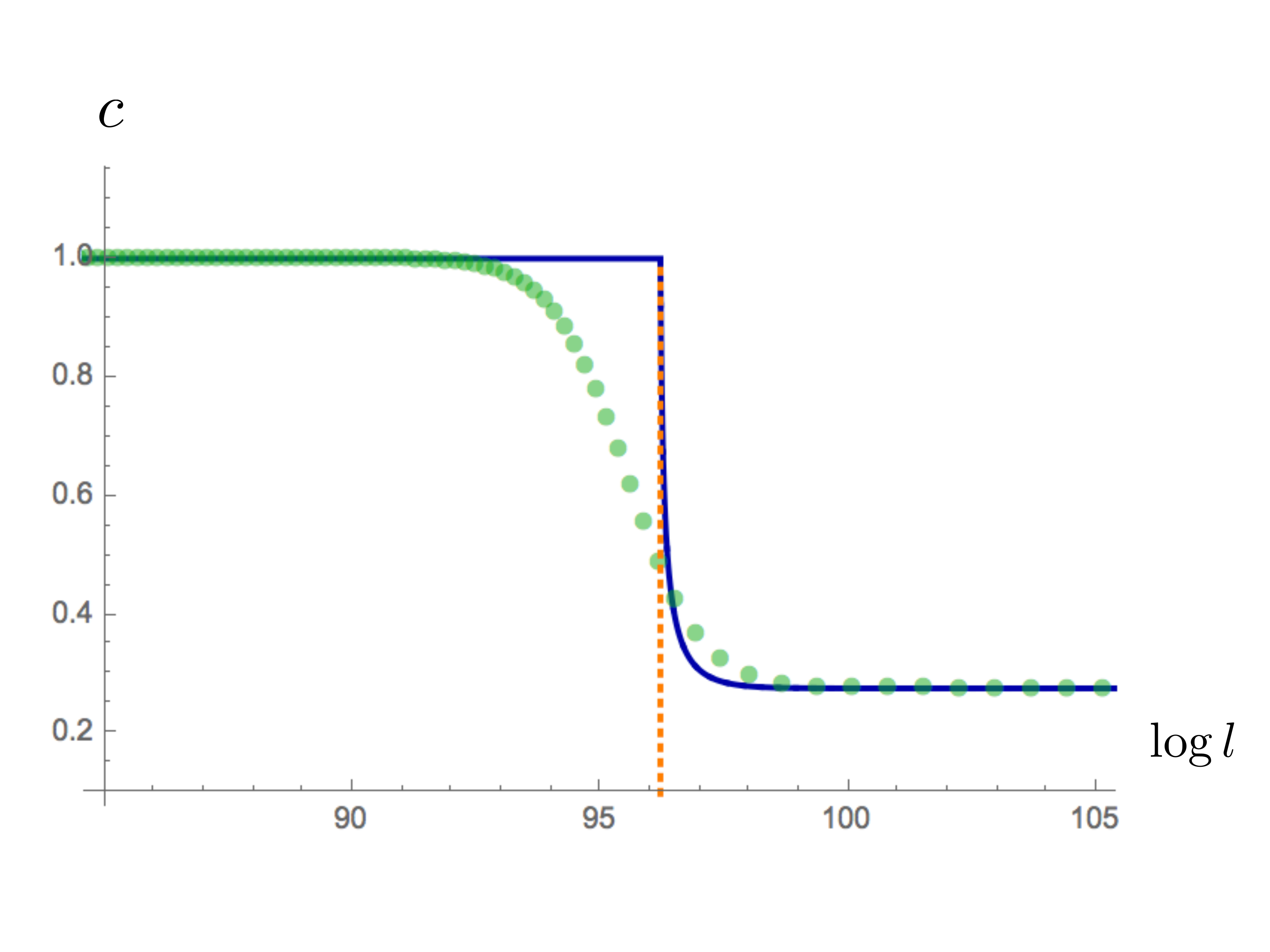}}
\subfigure[For $a^2 =0.98$]{\includegraphics[angle=0,width=0.49\textwidth]{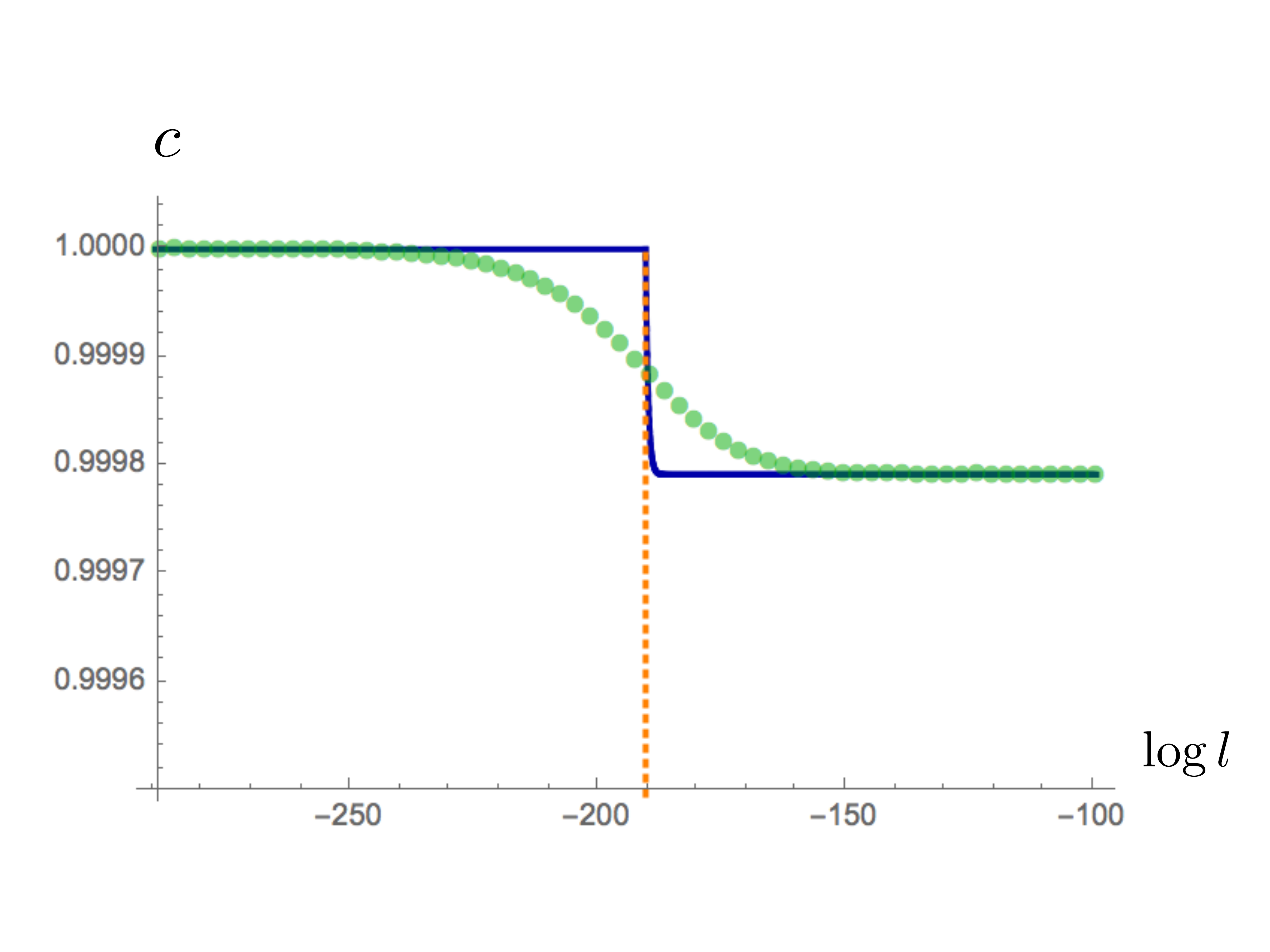}}
\caption{(color online) The entropic c-function evaluated in the $AdS_3^{uv}\rightarrow AdS_3^{ir}$ domain wall flows with different values of $a$. The reference central charge of $AdS_4^{uv}$, $c_{uv}=1$. The vertical dotted orange line is the position of the thin-wall, $y_{\text{DW}}=\log(2/\ell_{\text{DW}})$, where we assumed that $y_{\text{DW}}$ satisfies the conditions, $A'(y_{\text{DW}})<0$ and $A''(y_{\text{DW}})=0$. The $AdS_3^{ir}$ results agree excellently with the numerically computed values for large values of $l$.}
\label{cfunction}
\end{figure}

In Fig. \ref{cfunction}, we plot several  exact $c$-functions relying on the intrinsic parameter $a$. The result shows that the $c$-function always decreases monotonically regardless of the value of $a$. We checked that the $c$-function of the exact numerical result are well described by the thin-wall approximation only with some error near the wall.

%%%%%%%%%%%%%%%%%%%%%%%
%
%	Discussion
%
%%%%%%%%%%%%%%%%%%%%%%%

\section{Discussion}
\label{sec:7}

In this work, we have studied the RG flow of the entanglement entropy from a two-dimensional UV CFT to another two-dimensional IR CFT. On the dual gravity side, such a renormalization group flow of a quantum field theory was realized as a dual geometry interpolating two AdS geometries: one is an unstable AdS geometry at the asymptotic boundary, which corresponds to the UV CFT, and the other is an AdS space in the interior which describes a stable IR CFT. Following the holographic renormalization procedure, it was well known that the central charge of the dual CFT can be represented by the combination of the Newton constant and the AdS radius of the dual AdS space. In this work, we constructed a dual space interpolating two AdS spaces and showed that the holographic renormalization prescription exactly reproduces the known central charges of the dual CFT at two fixed points. Although the central charge is well defined in CFT, it is not easy work to define the $c$-function when the theory is deviated from CFT except the fixed point described by CFT. Applying the holographic renormalization prescription in this work, we defined the $c$-function reproducing the known central charges of UV and IR CFTs and also showed that such a $c$-function monotonically decreases along the renormalization group flow.

We further studied how the entanglement entropy flows along the renormalization group flow. The entanglement entropy was known as an important quantity which may play a central role in investigating a variety of quantum phases transitions. For an even-dimensional dual CFT, intriguingly, it was known that the logarithmic term of the entanglement entropy is associated with the central charge. In this work, we calculated the holographic entanglement entropy on a three-dimensional space interpolating two AdS spaces, which is dual to the relevant deformation of the UV CFT, and showed that the resulting entanglement entropy of the dual two-dimensional field theory exactly reproduces the expected entanglement entropy of a two-dimensional CFT at two UV and IR fixed points. In addition, we further showed that the $c$-function derived from the entanglement entropy reproduces the expected central charges of UV and IR CFTs and that it
satisfies the $c$-theorem. This result indicates that, when we perturb a two-dimensional CFT with a relevant deformation operator, the UV CFT smoothly flows to another CFT at IR without any phase transition. To see this feature analytically, we investigated a new thin-wall approximation in which we utilized a natural prescription different from the one used in \cite{Myers:2012ed}. The thin-wall approximation used in this work well approximates the qualitative features of the exact numerical calculation and showed that there is no phase transition breaking the $c$-theorem, which is consistent with the exact result.

\begin{figure}
\centering
\includegraphics[width=0.6\textwidth]{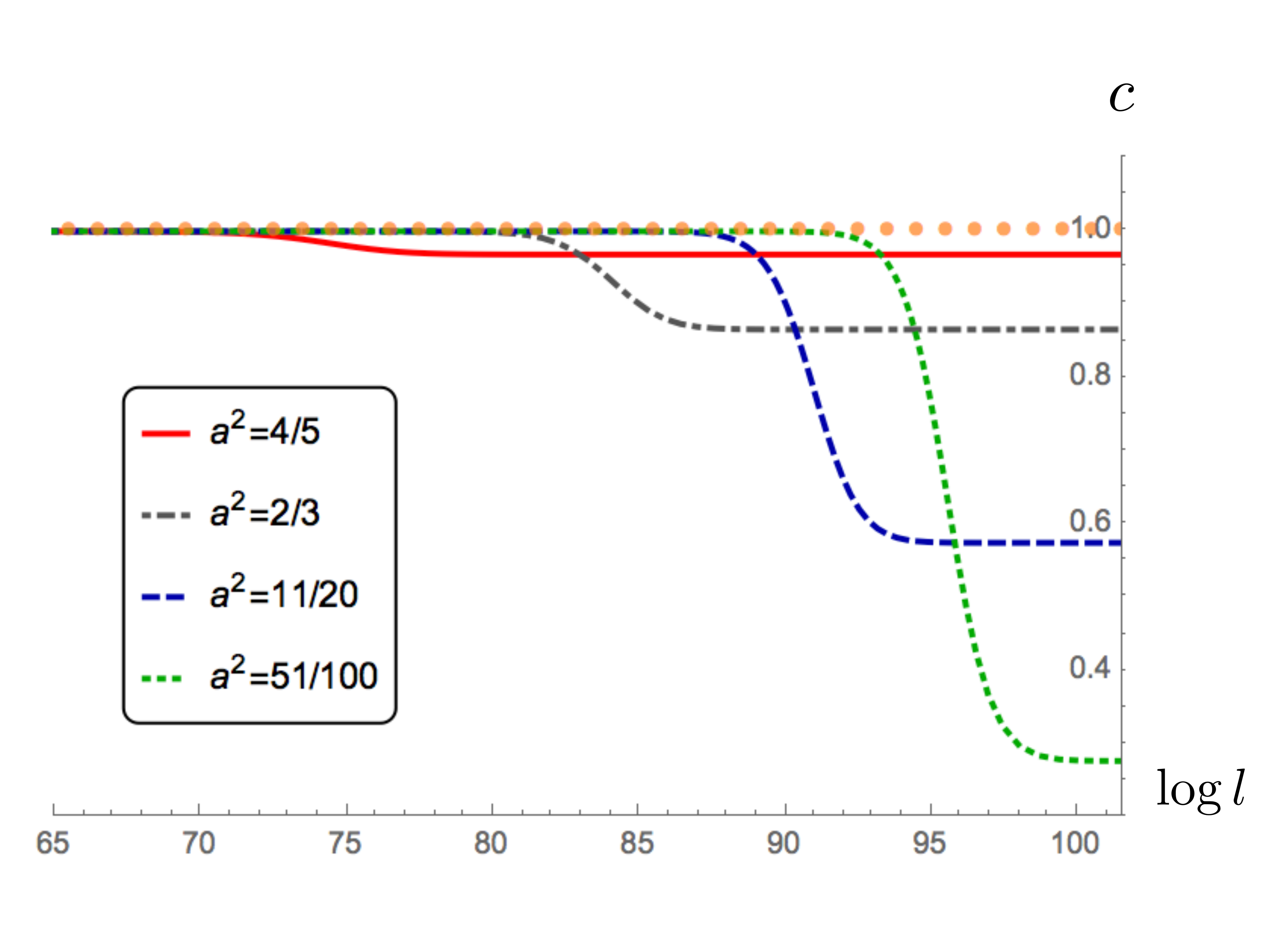}
\caption{(color online) The entropic c-function evaluated in the $AdS_3^{uv}\rightarrow AdS_3^{ir}$ domain wall flows with different values of $a$. The reference central charge of $AdS_4^{uv}$, $c_{uv}=1$, are shown as orange circle line. The $AdS_3^{ir}$ results agree excellently with the numerically computed values for large values of $l$.}
\label{cfunction}
\end{figure}

\vspace{0.5cm}

\section*{\small Acknowledgement}

This work was supported by the Korea Ministry of Education, Science and Technology, Gyeongsangbuk Do and Pohang City. JHL (NRF-2018R1A6A3A11049655), CP (NRF-2016R1D1A1B03932371), and DR (NRF-2017R1D1A1B03029430) were also supported by Basic Science Research Program through the National Research Foundation of Korea funded by the Ministry of Education.

%%%%%%%%%%%%%%%%%%%%%%%%%
%                                                                            %
%   The Bibliography                                             %
%                                                                            %
%%%%%%%%%%%%%%%%%%%%%%%%%
\providecommand{\href}[2]{#2}


\begin{thebibliography}{10}

%\cite{Calabrese:2004eu}
\bibitem{Calabrese:2004eu} 
  P.~Calabrese and J.~L.~Cardy,
  %``Entanglement entropy and quantum field theory,''
  J.\ Stat.\ Mech.\  {\bf 0406}, P06002 (2004)
  %doi:10.1088/1742-5468/2004/06/P06002
  [hep-th/0405152].
  %%CITATION = %doi:10.1088/1742-5468/2004/06/P06002;%%
  %786 citations counted in INSPIRE as of 10 Apr 2018

 %\cite{Calabrese:2005zw}
\bibitem{Calabrese:2005zw} 
  P.~Calabrese and J.~L.~Cardy,
  %``Entanglement entropy and quantum field theory: A Non-technical introduction,''
  Int.\ J.\ Quant.\ Inf.\  {\bf 4}, 429 (2006)
  %doi:10.1142/S021974990600192X
  [quant-ph/0505193].
  %%CITATION = %doi:10.1142/S021974990600192X;%%
  %133 citations counted in INSPIRE as of 10 Apr 2018 
  
  %\cite{Calabrese:2009qy}
\bibitem{Calabrese:2009qy} 
  P.~Calabrese and J.~Cardy,
  %``Entanglement entropy and conformal field theory,''
  J.\ Phys.\ A {\bf 42}, 504005 (2009)
  %doi:10.1088/1751-8113/42/50/504005
  [arXiv:0905.4013 [cond-mat.stat-mech]].
  %%CITATION = %doi:10.1088/1751-8113/42/50/504005;%%
  %402 citations counted in INSPIRE as of 10 Apr 2018
  
 %\cite{Maldacena:1997re}
\bibitem{Maldacena:1997re} 
  J.~M.~Maldacena,
  %``The Large N limit of superconformal field theories and supergravity,''
  Int.\ J.\ Theor.\ Phys.\  {\bf 38}, 1113 (1999)
  [Adv.\ Theor.\ Math.\ Phys.\  {\bf 2}, 231 (1998)]
  %%%doi:10.1023/A:1026654312961, 10.4310/ATMP.1998.v2.n2.a1
  [hep-th/9711200].
  %%CITATION = %%doi:10.1023/A:1026654312961, 10.4310/ATMP.1998.v2.n2.a1;%%
  %13652 citations counted in INSPIRE as of 04 May 2018

%\cite{Gubser:1998bc}
\bibitem{Gubser:1998bc} 
  S.~S.~Gubser, I.~R.~Klebanov and A.~M.~Polyakov,
  %``Gauge theory correlators from noncritical string theory,''
  Phys.\ Lett.\ B {\bf 428}, 105 (1998)
  %%%doi:10.1016/S0370-2693(98)00377-3
  [hep-th/9802109].
  %%CITATION = %%doi:10.1016/S0370-2693(98)00377-3;%%
  %7644 citations counted in INSPIRE as of 04 May 2018

%\cite{Witten:1998qj}
\bibitem{Witten:1998qj} 
  E.~Witten,
  %``Anti-de Sitter space and holography,''
  Adv.\ Theor.\ Math.\ Phys.\  {\bf 2}, 253 (1998)
  %%%doi:10.4310/ATMP.1998.v2.n2.a2
  [hep-th/9802150].
  %%CITATION = %%doi:10.4310/ATMP.1998.v2.n2.a2;%%
  %8912 citations counted in INSPIRE as of 04 May 2018 
  
  %\cite{,Witten:1998zw}
\bibitem{Witten:1998zw} 
  E.~Witten,
  %``Anti-de Sitter space, thermal phase transition, and confinement in gauge theories,''
  Adv.\ Theor.\ Math.\ Phys.\  {\bf 2}, 505 (1998)
  %doi:10.4310/ATMP.1998.v2.n3.a3
  [hep-th/9803131].
  %%CITATION = %doi:10.4310/ATMP.1998.v2.n3.a3;%%
  %2706 citations counted in INSPIRE as of 25 May 2018
  
  
  
  
  
  
  
  
  
  
  
  
  %\cite{Ryu:2006bv}
\bibitem{Ryu:2006bv}
  S.~Ryu and T.~Takayanagi,
  %``Holographic derivation of entanglement entropy from AdS/CFT,''
  Phys.\ Rev.\ Lett.\  {\bf 96}, 181602 (2006)
  %%%doi:10.1103/PhysRevLett.96.181602
  [hep-th/0603001].
  %%CITATION = %%doi:10.1103/PhysRevLett.96.181602;%%
  %1532 citations counted in INSPIRE as of 30 Apr 2018

%\cite{Ryu:2006ef}
\bibitem{Ryu:2006ef}
  S.~Ryu and T.~Takayanagi,
  %``Aspects of Holographic Entanglement Entropy,''
  JHEP {\bf 0608}, 045 (2006)
  %%%doi:10.1088/1126-6708/2006/08/045
  [hep-th/0605073].
  %%CITATION = %%doi:10.1088/1126-6708/2006/08/045;%%
  %923 citations counted in INSPIRE as of 30 Apr 2018
  
  %\cite{Hubeny:2007xt}
\bibitem{Hubeny:2007xt} 
  V.~E.~Hubeny, M.~Rangamani and T.~Takayanagi,
  %``A Covariant holographic entanglement entropy proposal,''
  JHEP {\bf 0707}, 062 (2007)
  %%doi:10.1088/1126-6708/2007/07/062
  [arXiv:0705.0016 [hep-th]].
  %%CITATION = %%doi:10.1088/1126-6708/2007/07/062;%%
  %607 citations counted in INSPIRE as of 15 May 2018
  
  %\cite{Casini:2011kv}
\bibitem{Casini:2011kv} 
  H.~Casini, M.~Huerta and R.~C.~Myers,
  %``Towards a derivation of holographic entanglement entropy,''
  JHEP {\bf 1105}, 036 (2011)
  %%doi:10.1007/JHEP05(2011)036
  [arXiv:1102.0440 [hep-th]].
  %%CITATION = %%doi:10.1007/JHEP05(2011)036;%%
  %565 citations counted in INSPIRE as of 15 May 2018
  
    %\cite{Lewkowycz:2013nqa}
\bibitem{Lewkowycz:2013nqa} 
  A.~Lewkowycz and J.~Maldacena,
  %``Generalized gravitational entropy,''
  JHEP {\bf 1308}, 090 (2013)
  %%doi:10.1007/JHEP08(2013)090
  [arXiv:1304.4926 [hep-th]].
  %%CITATION = %%doi:10.1007/JHEP08(2013)090;%%
  %376 citations counted in INSPIRE as of 15 May 2018

 %\cite{Casini:2004bw}
\bibitem{Casini:2004bw} 
  H.~Casini and M.~Huerta,
  %``A Finite entanglement entropy and the c-theorem,''
  Phys.\ Lett.\ B {\bf 600}, 142 (2004)
  %doi:10.1016/j.physletb.2004.08.072
  [hep-th/0405111].
  %%CITATION = %doi:10.1016/j.physletb.2004.08.072;%%
  %179 citations counted in INSPIRE as of 10 Apr 2018

 %\cite{Casini:2006es}
\bibitem{Casini:2006es} 
  H.~Casini and M.~Huerta,
  %``A c-theorem for the entanglement entropy,''
  J.\ Phys.\ A {\bf 40}, 7031 (2007)
  %doi:10.1088/1751-8113/40/25/S57
  [cond-mat/0610375].
  %%CITATION = %doi:10.1088/1751-8113/40/25/S57;%%
  %60 citations counted in INSPIRE as of 10 Apr 2018
  
%\cite{Albash:2011nq}
\bibitem{Albash:2011nq} 
  T.~Albash and C.~V.~Johnson,
  %``Holographic Entanglement Entropy and Renormalization Group Flow,''
  JHEP {\bf 1202}, 095 (2012)
  %%doi:10.1007/JHEP02(2012)095
  [arXiv:1110.1074 [hep-th]].
  %%CITATION = %doi:10.1007/JHEP02(2012)095;%%
  %43 citations counted in INSPIRE as of 13 Oct 2017
  
  %\cite{Klebanov:2012yf}
\bibitem{Klebanov:2012yf} 
  I.~R.~Klebanov, T.~Nishioka, S.~S.~Pufu and B.~R.~Safdi,
  %``On Shape Dependence and RG Flow of Entanglement Entropy,''
  JHEP {\bf 1207}, 001 (2012)
  %doi:10.1007/JHEP07(2012)001
  [arXiv:1204.4160 [hep-th]].
  %%CITATION = %doi:10.1007/JHEP07(2012)001;%%
  %54 citations counted in INSPIRE as of 25 May 2018
  
  %\cite{Cremonini:2013ipa}
\bibitem{Cremonini:2013ipa} 
  S.~Cremonini and X.~Dong,
  %``Constraints on renormalization group flows from holographic entanglement entropy,''
  Phys.\ Rev.\ D {\bf 89}, no. 6, 065041 (2014)
  %doi:10.1103/PhysRevD.89.065041
  [arXiv:1311.3307 [hep-th]].
  %%CITATION = %doi:10.1103/PhysRevD.89.065041;%%
  %10 citations counted in INSPIRE as of 25 May 2018
  
  %\cite{Faulkner:2014jva}
\bibitem{Faulkner:2014jva} 
  T.~Faulkner,
  %``Bulk Emergence and the RG Flow of Entanglement Entropy,''
  JHEP {\bf 1505}, 033 (2015)
  %doi:10.1007/JHEP05(2015)033
  [arXiv:1412.5648 [hep-th]].
  %%CITATION = %doi:10.1007/JHEP05(2015)033;%%
  %46 citations counted in INSPIRE as of 25 May 2018
 
   %\cite{Park:2014gja}
\bibitem{Park:2014gja} 
  C.~Park,
  %``Holographic renormalization in dense medium,''
  Adv.\ High Energy Phys.\  {\bf 2014}, 565219 (2014)
  %doi:10.1155/2014/565219
  [arXiv:1405.1490 [hep-th]].
  %%CITATION = %doi:10.1155/2014/565219;%%
  %8 citations counted in INSPIRE as of 23 May 2018
  
  %\cite{Park:2015hcz}
\bibitem{Park:2015hcz} 
  C.~Park,
  %``Thermodynamic law from the entanglement entropy bound,''
  Phys.\ Rev.\ D {\bf 93}, no. 8, 086003 (2016)
  %doi:10.1103/PhysRevD.93.086003
  [arXiv:1511.02288 [hep-th]].
  %%CITATION = %doi:10.1103/PhysRevD.93.086003;%%
  %6 citations counted in INSPIRE as of 23 May 2018
  
  %\cite{Casini:2015ffa}
\bibitem{Casini:2015ffa} 
  H.~Casini, E.~Teste and G.~Torroba,
  %``Holographic RG flows, entanglement entropy and the sum rule,''
  JHEP {\bf 1603}, 033 (2016)
  %doi:10.1007/JHEP03(2016)033
  [arXiv:1510.02103 [hep-th]].
  %%CITATION = %doi:10.1007/JHEP03(2016)033;%%
  %4 citations counted in INSPIRE as of 25 May 2018

  %\cite{Kim:2016ayz}
\bibitem{Kim:2016ayz} 
  K.~S.~Kim and C.~Park,
  %``Emergent geometry from field theory: Wilson?s renormalization group revisited,''
  Phys.\ Rev.\ D {\bf 93}, no. 12, 121702 (2016)
  %doi:10.1103/PhysRevD.93.121702
  [arXiv:1604.04990 [hep-th]].
  %%CITATION = %doi:10.1103/PhysRevD.93.121702;%%
  %4 citations counted in INSPIRE as of 23 May 2018
  
  %\cite{Kim:2016jwu}
\bibitem{Kim:2016jwu} 
  K.~S.~Kim and C.~Park,
  %``Renormalization group flow of entanglement entropy to thermal entropy,''
  Phys.\ Rev.\ D {\bf 95}, no. 10, 106007 (2017)
  %doi:10.1103/PhysRevD.95.106007
  [arXiv:1610.07266 [hep-th]].
  %%CITATION = %doi:10.1103/PhysRevD.95.106007;%%
  %2 citations counted in INSPIRE as of 23 May 201
   
  %\cite{Bueno:2016rma}
\bibitem{Bueno:2016rma} 
  P.~Bueno and W.~Witczak-Krempa,
  %``Holographic torus entanglement and its renormalization group flow,''
  Phys.\ Rev.\ D {\bf 95}, no. 6, 066007 (2017)
  %doi:10.1103/PhysRevD.95.066007
  [arXiv:1611.01846 [hep-th]].
  %%CITATION = %doi:10.1103/PhysRevD.95.066007;%%
  %4 citations counted in INSPIRE as of 25 May 2018
    
  %\cite{Kim:2016hig}
\bibitem{Kim:2016hig} 
  K.~S.~Kim, M.~Park, J.~Cho and C.~Park,
  %``Emergent geometric description for a topological phase transition in the Kitaev superconductor model,''
  Phys.\ Rev.\ D {\bf 96}, no. 8, 086015 (2017)
  %doi:10.1103/PhysRevD.96.086015
  [arXiv:1610.07312 [hep-th]].
  %%CITATION = %doi:10.1103/PhysRevD.96.086015;%%
  %3 citations counted in INSPIRE as of 23 May 2018
  
  %\cite{Kim:2017lyx}
\bibitem{Kim:2017lyx} 
  K.~S.~Kim, S.~B.~Chung and C.~Park,
  %``An emergent holographic description for the Kondo effect: The role of an extra dimension in a non-perturbative field theoretical approach,''
  arXiv:1705.06571 [hep-th].
  %%CITATION = ARXIV:1705.06571;%%
  %2 citations counted in INSPIRE as of 23 May 2018
  
  %\cite{Jang:2017gwd}
\bibitem{Jang:2017gwd} 
  O.~K.~Kwon, D.~Jang, Y.~Kim and D.~D.~Tolla,
  %``Gravity from Entanglement and RG Flow in a Top-down Approach,''
  JHEP {\bf 1805}, 009 (2018)
  %doi:10.1007/JHEP05(2018)009
  [arXiv:1712.09101 [hep-th]].
  %%CITATION = %doi:10.1007/JHEP05(2018)009;%%
  %2 citations counted in INSPIRE as of 25 May 2018
  
 %\cite{Ghosh:2017big}
\bibitem{Ghosh:2017big} 
  J.~K.~Ghosh, E.~Kiritsis, F.~Nitti and L.~T.~Witkowski,
  %``Holographic RG flows on curved manifolds and quantum phase transitions,''
  arXiv:1711.08462 [hep-th].
  %%CITATION = ARXIV:1711.08462;%%
  
  %\cite{Narayanan:2018ilr}
\bibitem{Narayanan:2018ilr} 
  R.~Narayanan, C.~Park and Y.~L.~Zhang,
  %``Entanglement Entropy of Randomly Disordered System,''
  arXiv:1803.01064 [hep-th].
  %%CITATION = ARXIV:1803.01064;%%
  
   %\cite{Koh:2018rsw}
\bibitem{Koh:2018rsw} 
  S.~Koh, J.~Hun Lee, C.~Park and D.~Ro,
  %``Quantum entanglement in inflationary cosmology,''
  arXiv:1806.01092 [hep-th].
  %%CITATION = ARXIV:1806.01092;%%
  
  %\cite{Deger:1999st}
\bibitem{Deger:1999st} 
  N.~S.~Deger, A.~Kaya, E.~Sezgin and P.~Sundell,
  %``Matter coupled AdS(3) supergravities and their black strings,''
  Nucl.\ Phys.\ B {\bf 573}, 275 (2000)
  %%doi:10.1016/S0550-3213(99)00734-8
  [hep-th/9908089].
  %%CITATION = %doi:10.1016/S0550-3213(99)00734-8;%%
  %32 citations counted in INSPIRE as of 02 Aug 2017
  
  %\cite{Deger:2002hv}
\bibitem{Deger:2002hv} 
  N.~S.~Deger,
  %``Renormalization group flows from D = 3, N=2 matter coupled gauged supergravities,''
  JHEP {\bf 0211}, 025 (2002)
  %%doi:10.1088/1126-6708/2002/11/025
  [hep-th/0209188].
  %%CITATION = %doi:10.1088/1126-6708/2002/11/025;%%
  %13 citations counted in INSPIRE as of 01 Aug 2017

  
%\cite{Zamolodchikov:1986gt}
\bibitem{Zamolodchikov:1986gt} 
  A.~B.~Zamolodchikov,
  %``Irreversibility of the Flux of the Renormalization Group in a 2D Field Theory,''
  JETP Lett.\  {\bf 43}, 730 (1986)
  [Pisma Zh.\ Eksp.\ Teor.\ Fiz.\  {\bf 43}, 565 (1986)].
  %%CITATION = JTPLA,43,730;%%
  %1233 citations counted in INSPIRE as of 10 Apr 2018

%\cite{Cardy:1988cwa}
\bibitem{Cardy:1988cwa} 
  J.~L.~Cardy,
  %``Is There a c Theorem in Four-Dimensions?,''
  Phys.\ Lett.\ B {\bf 215}, 749 (1988).
 % %doi:10.1016/0370-2693(88)90054-8
  %%CITATION = %doi:10.1016/0370-2693(88)90054-8;%%
  %365 citations counted in INSPIRE as of 10 Apr 2018

%\cite{Komargodski:2011vj}
\bibitem{Komargodski:2011vj} 
  Z.~Komargodski and A.~Schwimmer,
  %``On Renormalization Group Flows in Four Dimensions,''
  JHEP {\bf 1112}, 099 (2011)
  %doi:10.1007/JHEP12(2011)099
  [arXiv:1107.3987 [hep-th]].
  %%CITATION = %doi:10.1007/JHEP12(2011)099;%%
  %374 citations counted in INSPIRE as of 10 Apr 2018
  
  %\cite{Komargodski:2011xv}
\bibitem{Komargodski:2011xv} 
  Z.~Komargodski,
  %``The Constraints of Conformal Symmetry on RG Flows,''
  JHEP {\bf 1207}, 069 (2012)
  %doi:10.1007/JHEP07(2012)069
  [arXiv:1112.4538 [hep-th]].
  %%CITATION = %doi:10.1007/JHEP07(2012)069;%%
  %166 citations counted in INSPIRE as of 10 Apr 2018

%\cite{Myers:2012ed}
\bibitem{Myers:2012ed} 
  R.~C.~Myers and A.~Singh,
  %``Comments on Holographic Entanglement Entropy and RG Flows,''
  JHEP {\bf 1204}, 122 (2012)
  %doi:10.1007/JHEP04(2012)122
  [arXiv:1202.2068 [hep-th]].
  %%CITATION = %doi:10.1007/JHEP04(2012)122;%%
  %93 citations counted in INSPIRE as of 25 May 2018

%\cite{Gubser:2000nd}
\bibitem{Gubser:2000nd} 
  S.~S.~Gubser,
  %``Curvature singularities: The Good, the bad, and the naked,''
  Adv.\ Theor.\ Math.\ Phys.\  {\bf 4}, 679 (2000)
  %doi:10.4310/ATMP.2000.v4.n3.a6
  [hep-th/0002160].
  %%CITATION = %doi:10.4310/ATMP.2000.v4.n3.a6;%%
  %350 citations counted in INSPIRE as of 25 May 2018
    
  %\cite{Liu:2012eea}
\bibitem{Liu:2012eea} 
  H.~Liu and M.~Mezei,
  %``A Refinement of entanglement entropy and the number of degrees of freedom,''
  JHEP {\bf 1304}, 162 (2013)
  %doi:10.1007/JHEP04(2013)162
  [arXiv:1202.2070 [hep-th]].
  %%CITATION = %doi:10.1007/JHEP04(2013)162;%%
  %153 citations counted in INSPIRE as of 23 May 2018
  
  %\cite{Kim:2014yca}
\bibitem{Kim:2014yca} 
  K.~K.~Kim, O.~K.~Kwon, C.~Park and H.~Shin,
  %``Renormalized Entanglement Entropy Flow in Mass-deformed ABJM Theory,''
  Phys.\ Rev.\ D {\bf 90}, no. 4, 046006 (2014)
  %doi:10.1103/PhysRevD.90.046006
  [arXiv:1404.1044 [hep-th]].
  %%CITATION = %doi:10.1103/PhysRevD.90.046006;%%
  %18 citations counted in INSPIRE as of 23 May 2018
  
  %\cite{Kim:2014qpa}
\bibitem{Kim:2014qpa} 
  K.~K.~Kim, O.~K.~Kwon, C.~Park and H.~Shin,
  %``Holographic entanglement entropy of mass-deformed Aharony-Bergman-Jafferis-Maldacena theory,''
  Phys.\ Rev.\ D {\bf 90}, no. 12, 126003 (2014)
  %doi:10.1103/PhysRevD.90.126003
  [arXiv:1407.6511 [hep-th]].
  %%CITATION = %doi:10.1103/PhysRevD.90.126003;%%
  %14 citations counted in INSPIRE as of 23 May 2018
  
 %\cite{Park:2015afa}
\bibitem{Park:2015afa} 
  C.~Park,
  %``Holographic entanglement entropy in the nonconformal medium,''
  Phys.\ Rev.\ D {\bf 91}, no. 12, 126003 (2015)
  %doi:10.1103/PhysRevD.91.126003
  [arXiv:1501.02908 [hep-th]].
  %%CITATION = %doi:10.1103/PhysRevD.91.126003;%%
  %13 citations counted in INSPIRE as of 23 May 2018 
  
  %\cite{Park:2015dia}
\bibitem{Park:2015dia} 
  C.~Park,
  %``Logarithmic Corrections to the Entanglement Entropy,''
  Phys.\ Rev.\ D {\bf 92}, no. 12, 126013 (2015)
  %doi:10.1103/PhysRevD.92.126013
  [arXiv:1505.03951 [hep-th]].
  %%CITATION = %doi:10.1103/PhysRevD.92.126013;%%
  %10 citations counted in INSPIRE as of 23 May 2018
  
  %\cite{Kim:2018mgz}
\bibitem{Kim:2018mgz} 
  K.~K.~Kim, C.~Park, J.~Hun Lee and B.~Ahn,
  %``Holographic Entanglement Entropy with Momentum Relaxation,''
  arXiv:1804.00412 [hep-th].
  %%CITATION = ARXIV:1804.00412;%%
  
 %\cite{Henningson:1998gx}
\bibitem{Henningson:1998gx} 
  M.~Henningson and K.~Skenderis,
  %``The Holographic Weyl anomaly,''
  JHEP {\bf 9807}, 023 (1998)
 % doi:10.1088/1126-6708/1998/07/023
  [hep-th/9806087].
  %%CITATION = doi:10.1088/1126-6708/1998/07/023;%%
  %1219 citations counted in INSPIRE as of 19 Jun 2018
  
 %\cite{Brown:1986nw}
\bibitem{Brown:1986nw} 
  J.~D.~Brown and M.~Henneaux,
  %``Central Charges in the Canonical Realization of Asymptotic Symmetries: An Example from Three-Dimensional Gravity,''
  Commun.\ Math.\ Phys.\  {\bf 104}, 207 (1986).
  %doi:10.1007/BF01211590
  %%CITATION = doi:10.1007/BF01211590;%%
  %1535 citations counted in INSPIRE as of 19 Jun 2018 
  
%\cite{Barnes:2004jj}
\bibitem{Barnes:2004jj} 
  E.~Barnes, K.~A.~Intriligator, B.~Wecht and J.~Wright,
  %``Evidence for the strongest version of the 4d a-theorem, via a-maximization along RG flows,''
  Nucl.\ Phys.\ B {\bf 702}, 131 (2004)
 % doi:10.1016/j.nuclphysb.2004.09.016
  [hep-th/0408156].
  %%CITATION = doi:10.1016/j.nuclphysb.2004.09.016;%%
  %96 citations counted in INSPIRE as of 19 Jun 2018
  

%\cite{Kim:2015rvu}
\bibitem{Kim:2015rvu} 
  N.~Kim and J.~Hun Lee,
  %``Time-evolution of the holographic entanglement entropy and metric perturbationst,''
  J.\ Korean Phys.\ Soc.\  {\bf 69}, no. 4, 623 (2016)
 % doi:10.3938/jkps.69.623
  [arXiv:1512.02816 [hep-th]].
  %%CITATION = doi:10.3938/jkps.69.623;%%
  %5 citations counted in INSPIRE as of 19 Jun 2018  


%\cite{Skenderis:1999mm}
\bibitem{Skenderis:1999mm} 
  K.~Skenderis and P.~K.~Townsend,
  %``Gravitational stability and renormalization group flow,''
  Phys.\ Lett.\ B {\bf 468}, 46 (1999)
  %doi:10.1016/S0370-2693(99)01212-5
  [hep-th/9909070].
  %%CITATION = doi:10.1016/S0370-2693(99)01212-5;%%
  %354 citations counted in INSPIRE as of 19 Jun 2018
  






%%%%%%%%%%%%%%%%%%%%%%%%%%%%%%%%%%%%%%%

\end{thebibliography}
\end{document}